\pdfoutput=1
\documentclass[astrosymb,tighten,preprint2]{aastex63}

\usepackage[T1]{fontenc}
\usepackage[utf8]{inputenc}
\usepackage[icelandic,english]{babel}
\usepackage{gensymb}
\usepackage{lmodern}
\usepackage{hyperref}
\usepackage{amsmath}
\usepackage{natbib}
\setcitestyle{citesep={;}}
\setcitestyle{aysep={}}

\shorttitle{The Epoch of Giant Planet Migration Planet Search Program.}
\shortauthors{Tran et al.}
\graphicspath{{./}{Figures/}}

\begin{document}

\title{The Epoch of Giant Planet Migration Planet Search Program. I. \\
Near-Infrared Radial Velocity Jitter of Young Sun-like Stars}

\correspondingauthor{Quang H. Tran}
\author[0000-0001-6532-6755]{Quang H. Tran}
\email{quangtran@utexas.edu}
\affiliation{Department of Astronomy, The University of Texas at Austin, 2515 Speedway, Stop C1400, Austin, TX 78712, USA}

\author[0000-0003-2649-2288]{Brendan P. Bowler}
\affiliation{Department of Astronomy, The University of Texas at Austin, 2515 Speedway, Stop C1400, Austin, TX 78712, USA}

\author[0000-0001-9662-3496]{William D. Cochran}
\affiliation{Department of Astronomy, The University of Texas at Austin, 2515 Speedway, Stop C1400, Austin, TX 78712, USA}
\affiliation{McDonald Observatory, The University of Texas at Austin, 2515 Speedway, Stop C1400, Austin, TX 78712, USA}

\author{Michael Endl}
\affiliation{Department of Astronomy, The University of Texas at Austin, 2515 Speedway, Stop C1400, Austin, TX 78712, USA}
\affiliation{McDonald Observatory, The University of Texas at Austin, 2515 Speedway, Stop C1400, Austin, TX 78712, USA}

\author[0000-0001-7409-5688]{Guðmundur Stef{\'a}nsson}
\affiliation{Department of Astrophysical Sciences, Princeton University, 4 Ivy Lane, Princeton, NJ 08540, USA}
\affiliation{Department of Astronomy \& Astrophysics, The Pennsylvania State University, 525 Davey Lab, University Park, PA 16802, USA}
\affiliation{Center for Exoplanets \& Habitable Worlds, University Park, PA 16802, USA}
\affiliation{NASA Earth and Space Science Fellow}
\affiliation{Henry Norris Russell Fellow}

\author[0000-0001-9596-7983]{Suvrath Mahadevan}
\affiliation{Department of Astronomy \& Astrophysics, The Pennsylvania State University, 525 Davey Lab, University Park, PA 16802, USA}
\affiliation{Center for Exoplanets \& Habitable Worlds, University Park, PA 16802, USA}
\affiliation{Penn State Astrobiology Research Center, University Park, PA 16802, USA}

\author[0000-0001-8720-5612]{Joe P. Ninan}
\affiliation{Department of Astronomy \& Astrophysics, The Pennsylvania State University, 525 Davey Lab, University Park, PA 16802, USA}
\affiliation{Center for Exoplanets \& Habitable Worlds, University Park, PA 16802, USA}

\author[0000-0003-4384-7220]{Chad F. Bender}
\affiliation{Steward Observatory, University of Arizona, 933 N Cherry Ave, Tucson, AZ 85721, USA}

\author[0000-0003-1312-9391]{Samuel Halverson}
\affiliation{Jet Propulsion Laboratory, California Institute of Technology, 4800 Oak Grove Drive, Pasadena, CA 91109, USA}

\author[0000-0001-8127-5775]{Arpita Roy}
\affiliation{The Space Telescope Science Institute, 3700 San Martin Drive, Baltimore, MD 21218, USA}

\author[0000-0002-4788-8858]{Ryan C. Terrien}
\affiliation{Carleton College, One North College St., Northfield, MN 55057, USA}

\begin{abstract}
	We present early results from the Epoch of Giant Planet Migration program, a precise RV survey of over one hundred intermediate-age ($\sim$20--200~Myr) G and K dwarfs with the Habitable--Zone Planet Finder spectrograph (HPF) at McDonald Observatory's Hobby-Eberly Telescope (HET). The goals of this program are to determine the timescale and dominant physical mechanism of giant planet migration interior to the water ice line of Sun-like stars. Here, we summarize results from the first 14 months of this program, with a focus on our custom RV pipeline for HPF, a measurement of the intrinsic near-infrared RV activity of young Solar analogs, and modeling the underlying population-level distribution of stellar jitter. We demonstrate on-sky stability at the sub-2~m~s$^{-1}$ level for the K2 standard HD 3765 using a least-squares matching method to extract precise RVs. Based on a subsample of 29 stars with at least three RV measurements from our program, we find a median RMS level of 34~m~s$^{-1}$. This is nearly a factor of 2 lower than the median RMS level in the optical of 60~m~s$^{-1}$ for a comparison sample with similar ages and spectral types as our targets. The observed near-infrared jitter measurements for this subsample are well reproduced with a log-normal parent distribution with $\mu=4.15$ and $\sigma=1.02$. Finally, by compiling RMS values from previous planet search programs, we show that near-infrared jitter for G and K dwarfs generally decays with age in a similar fashion to optical wavelengths, albeit with a shallower slope and lower overall values for ages $\lesssim$1~Gyr.

\end{abstract}

\keywords{Instrumentation: precision spectrographs -- Planetary systems -- Stars: activity -- Techniques: radial velocities}

\section{\textbf{Introduction}}

\label{sec:intro}
    Beginning with the discovery of the first exoplanet around a main sequence star, 51 Peg b (\citealt{51Peg1995}), giant planets have received a disproportionate amount of attention relative to their intrinsic occurrence rate. This is because giant planets are generally the most easily detectable population of exoplanets: they induce large radial velocity (RV) amplitudes on their host stars and produce deep transit depths in comparison to the more prevalent Super-Earths and Neptunes. However, despite the continually growing number of Hot Jupiters and warm Jupiters\footnote{Following \citet{Gaudi2005} and \citet{Huang2016}, we define hot Jupiters as having $P \lesssim 10$ d and warm Jupiters as having $P \sim10-200$ d.} orbiting relatively old field stars ($\sim$1--10 Gyr), the origin of gas giants interior to the water ice line---located at $\approx$2.5 AU for a Sun-like star \citep{Martin2012}---remains unclear.
    
    Most giant planets are expected to have formed beyond the water ice line where their assembly is most efficient. Outside of this boundary, temperatures are low enough for volatile molecules to condense into ices which facilitates the rapid growth of cores \citep{Alibert2005, Albrecht2012}. During this process, a giant planet core must grow sufficiently massive to initiate runaway gas accretion before the dispersal of the gas disk. If there is enough material and the accretion timescale is faster than the disk dispersal timescale, then a giant planet may form \citep{Pollack1996}. If neither condition is met then the final planet mass and core mass fraction will be truncated. The ice line represents a critical barrier that facilitates this process to occur efficiently and thus sets the initial orbital conditions for most giant planets.
    
    Contrary to these predictions, there exists an appreciable population of gas giants interior to the water ice line. \citet{Johnson2010} measured a frequency of $6.5 \pm 0.7$\% for planets with masses $>$0.5 $M_\mathrm{Jup}$ orbiting within 2.5 AU of old (several Gyr) Sun-like stars. This frequency rises to about 15\% around metal-rich ([Fe/H] $> 0.25$) stars. \citet{Wittenmyer2020} measured a similar frequency of about $6.2^{+2.8}_{-1.6}$\% around F, G, and K stars at wider separations between 3--7 AU. This suggests that roughly half of giant planets orbiting Sun-like stars within $\approx$7 AU are located interior to the water ice line. If they formed exterior to this boundary, as expected, then they likely experienced inward orbital migration within the first few Gyr after their formation. Identifying the dominant migration pathway and its characteristic timescale has emerged as a major open problem in planet formation and evolution.
    
    Several theories have been developed to explain this population. One scenario for the origin of these close-in giant planets is \textit{in-situ} formation \citep{Batygin2016, Boley2016, Huang2016}. However, this formation route alone cannot explain observations of stellar spin-planet orbit misalignments \citep[e.g.,][]{Huber2013}, WJs on highly eccentric orbits \citep[e.g.,][]{Winn2015}, and the general dearth of additional low-mass planets in systems hosting close-in giant planets \citep[e.g.,][]{Knutson2014}, all suggesting a more violent, dynamic migration pathway. Instead, two general processes can best explain this population of giant planets: inspiraling disk migration or three-body dynamical interactions with an outer companion---for example, planet-planet scattering or high-eccentricity tidal migration \citep[e.g.,][]{Ward1997, Wu2003, Fabrycky2007, Juric2008, Chatterjee2008, Triaud2010, Naoz2011, Kley2012, Albrecht2012, Batygin2012}. These mechanisms predict a broad range of migration timescales and system architectures. However, none alone is able to explain the ensemble of observational properties of the giant planet population including their period-eccentricity distribution, occurrence rates, mass-period distribution, spin-orbit geometry, multi-planet architectures, and wide binary fraction \citep[e.g.,][]{Levrard2009,Huber2013, Fabrycky2014, Morton2014, Ngo2015, Mazeh2015, Pu2015, Wang2015a, Bryan2016, Mustill2017}.
    
    Fortunately, these migration pathways can be observationally distinguished by measuring the frequency of gas giants over time because they operate on different characteristic timescales. Disk migration must have taken place prior to $\sim$10 Myr, before the gas disk disappeared, whereas dynamical mechanisms occur on the order of $10^7$--$10^{10}$ yrs depending on the initial properties, orbital architecture, and dynamical history of the system \citep{Dawson2018}. Studying planetary systems over a wide range of ages can identify the characteristic timescale and dominant channel of planet migration.

    Young stars are intrinsically more magnetically active than older stars and induce non-dynamical astrophysical variations, which can mask RV signals from planets and are challenging to both identify and mitigate. These many contributions include magnetically induced active regions such as faculae, plages, and starspots; net magnetic suppression of convective blueshift \citep{Lagrange2010, Dumusque2011b, Boisse2012}; and long term magnetic cycles \citep{Dravins1982, Santos2010, Meunier2013}. Additional stellar-based RV variations originate from granulation \citep{DelMoro2004, Gray2009, Dumusque2011a}; acoustic pressure (p-mode) pulsations \citep{Kjeldsen1995, Aerts2019};
    Doppler broadening of absorption lines \citep{Gray1999}; and rotationally modulated surface features \citep{Vanderburg2016}.
    
    Among these, starspot-induced variability is especially pernicious. Starspots arise due to strong, buckled magnetic field lines that are organized into flux tubes and extend beyond the stellar surface. These magnetic tubes inhibit convection in localized regions when they intersect the stellar photosphere. The net result is a local suppression of the convective blueshift and a reduction in temperature; spots are cooler and redshifted compared to the surrounding photosphere. If the region is large enough, it appears darker and will influence the spectral line profile shape as the spot (or family of spots) rotates across the stellar disk \citep{Jahn1984, Donati1992, Schuessler1996, Fischer2016}. These effects are exacerbated at young ages when rotation is faster, magnetic fields are stronger, and spot surface covering fractions are higher.
    
    Granulation, convection, pulsations, and large rotationally-modulated starspots operate on a range of timescales spanning minutes to days and can contribute correlated signals that mimic the periodic RV variation of a planet. At optical wavelengths, jitter levels scale approximately logarithmically with age. At 5 Myr, for example, jitter can reach hundreds of meters per second, completely obscuring the signals of even the closest, most massive HJs. Even at 50 Myr, jitter can contribute RV signals as high as $\sim$100 m s$^{-1}$ \citep{Hillenbrand2015}. Thus, the \{130, 40, 30, 20\} m s$^{-1}$ RV semi-amplitudes created by a 1 $M_\mathrm{Jup}$ planet at separations of \{0.05, 0.5, 1, 2\} AU around a Solar-mass star are difficult to detect with precision RV optical spectrographs \citep{Donati2016, Johns-Krull2016, Yu2017}.
    
    Fortunately, by moving from the optical into the near-infrared (NIR), RV semi-amplitudes due to jitter have been shown to be reduced by a factor of $\approx$2--3 for active, young T Tauri stars ($\sim$1--10 Myr) and somewhat older members of young moving groups (20-150 Myr) \citep{Crockett2012, Bailey2012, Gagne2016}. At NIR wavelengths, the spectral energy distributions of Sun-like stars reach the Rayleigh--Jeans tail which reduces the contrast between starspots and the stellar photosphere; as a result, the effect of the spots on the absorption line profile is mitigated and activity-induced jitter is reduced.

    In this study, we present the first results from a new near-infrared high-precision RV survey of nearby young stars. The structure of this paper is as follows. In Section \ref{sec:Workingtitle}, we describe our survey and the primary science objectives. Our approach to measuring precise RVs and activity indicators is detailed in Section \ref{sec:Pipeline}. Our analysis of near-infrared stellar activity at intermediate ages is reported in Section \ref{sec:Jitter}. In Section \ref{sec:Dicussion}, we discuss how our findings fit with results from other RV surveys in optical and NIR wavelengths. Finally, we summarize our conclusions in Section \ref{sec:Summary}.
    
\section{\textbf{Establishing the Epoch of Giant Planet Migration}}\label{sec:Workingtitle}

\subsection{A Near-Infrared Precision Radial Velocity Survey of Young Stars}
    
    The recently commissioned Habitable--Zone Planet Finder (HPF) spectrograph (PI: S. Mahadevan) at McDonald Observatory’s 9.2-meter Hobby-Eberly Telescope (HET) is an ideal instrument to leverage this wavelength dependence to detect giant planets around intermediate-age stars. HPF is a near-infrared (0.81-1.27 $\mu$m), fiber fed, and environmentally stabilized high-resolution echelle spectrograph that is wavelength calibrated using a laser frequency comb (LFC) in order to achieve precise RVs at the m s$^{-1}$ level \citep{Mahadevan2012, Stefansson2016}. HPF has recently demonstrated 1.5 m s$^{-1}$ (RMS) nightly binned RV precision for Barnard’s star (M4 dwarf, $J$=5.2 mag) based on 118 high signal-to-noise (S/N) measurements \citep{Metcalf2019a}. This is a comparable precision in the near-infrared to optical high-resolution spectrographs such as CARMENES \citep[$\sim$1$-2$ m s$^{-1}$;][]{Quirrenbach2016, Tal-Or2018}, HARPS \citep[$\sim$1 m s$^{-1}$;][]{Guillem2012}, HIRES \citep[$1-2$ m s$^{-1}$;][]{Butler2017}, Magellan/PFS \citep[$\sim$1 m s$^{-1}$;][]{Crane2010}, and MINERVA \citep[1.8 m s$^{-1}$;][]{Wilson2019}.
    
    The Epoch of Giant Planet Migration planet search program was launched in 2018 to constrain the dominant inward migration route of giant planets by measuring their occurrence rate around intermediate-age ($\sim$20--200 Myr) Sun-like (G and K dwarfs) stars with HPF. Moving to somewhat older ages compared to previous RV surveys of T Tauri stars further reduces activity and opens the possibility of consistently reaching close-in young planets with velocity semi-amplitudes of tens to hundreds of m s$^{-1}$. The goal of this four-year survey is to study the entire population of giant planets interior to the water ice line ($a < 2.5$ AU) around young stars.  
    
    We have designed this survey to be directly comparable to similar RV programs that targeted field ($>$1 Gyr) stars \citep[e.g.,][]{Johnson2010}. If we measure a lower occurrence rate, then a dynamical origin is the dominant migration channel, since these processes take as long or longer ($\sim$10$^{7}$--10$^{10}$ yrs) to operate than the stellar ages ($\sim$20--200 Myr). Alternatively, if the occurrence rate is statistically similar to the field rate then this would indicate that the migration pathway must be rapid ($\lesssim$200 Myr), supporting disk migration or very early planet scattering. If we find a statistically higher occurrence rate, this would suggest that early disk migration and planet engulfment via tidal decay by the host star are common \citep[e.g.,][]{Rasio1996, Jackson2009, Hebb2010, Patra2017, Yee2020}.
    
    Our targets primarily draw from young stars in well-characterized, nearby young moving groups (YMGs), which are kinematically comoving clusters of stars in the solar neighborhood ($\lesssim$150 pc). Due to their proximity, brightness, and youth, YMG members are excellent targets for finding and characterizing young exoplanets. Targets are selected from compilations of established members \citep[e.g.,][]{Stauffer2007, Torres2008, Malo2013, Bell2015, Gagne2018a}. Our sample comprises bona fide and high-probability candidate members of ten YMGs---AB Dor, $\beta$ Pic, Carina, Carina-Near, Octans, Tuc-Hor, Pleiades, 32 Ori, Argus, and Pisces---and span a relatively narrow range of ages between 20 and 200 Myr. Ages for YMGs are usually derived from isochronal models using known members and have typical uncertainties of 10-20\%. To make efficient use of HPF time, all targets have been crossed-matched with Gaia DR2 and the Washington Double Star Catalog \citep{Gaia2018, Mason2019} to identify and remove known and suspected binaries with projected separations \textless1000 AU. This cut was selected so that Kozai-Lidov oscillations from any remaining wide stellar companions would operate on negligible ($>$ stellar age) timescales. Declination cuts were applied to account for the visibility of the HET, which sits at a fixed elevation angle of 55$^\circ$ and has access to a ring of declinations between $-11^\circ$ and +71$^\circ$ at a given time. Finally, all fast rotators with known $v$sin$i$ values $>$30 km s$^{-1}$ have been removed. This threshold was chosen to limit astrophysical jitter and mitigate reduced RV measurement precision for fast rotators, as stellar activity saturates at 30 km s$^{-1}$ \citep{White2007}.
    
\begin{figure*}[!t]
	    \centerline{\includegraphics[width=1.0\linewidth]{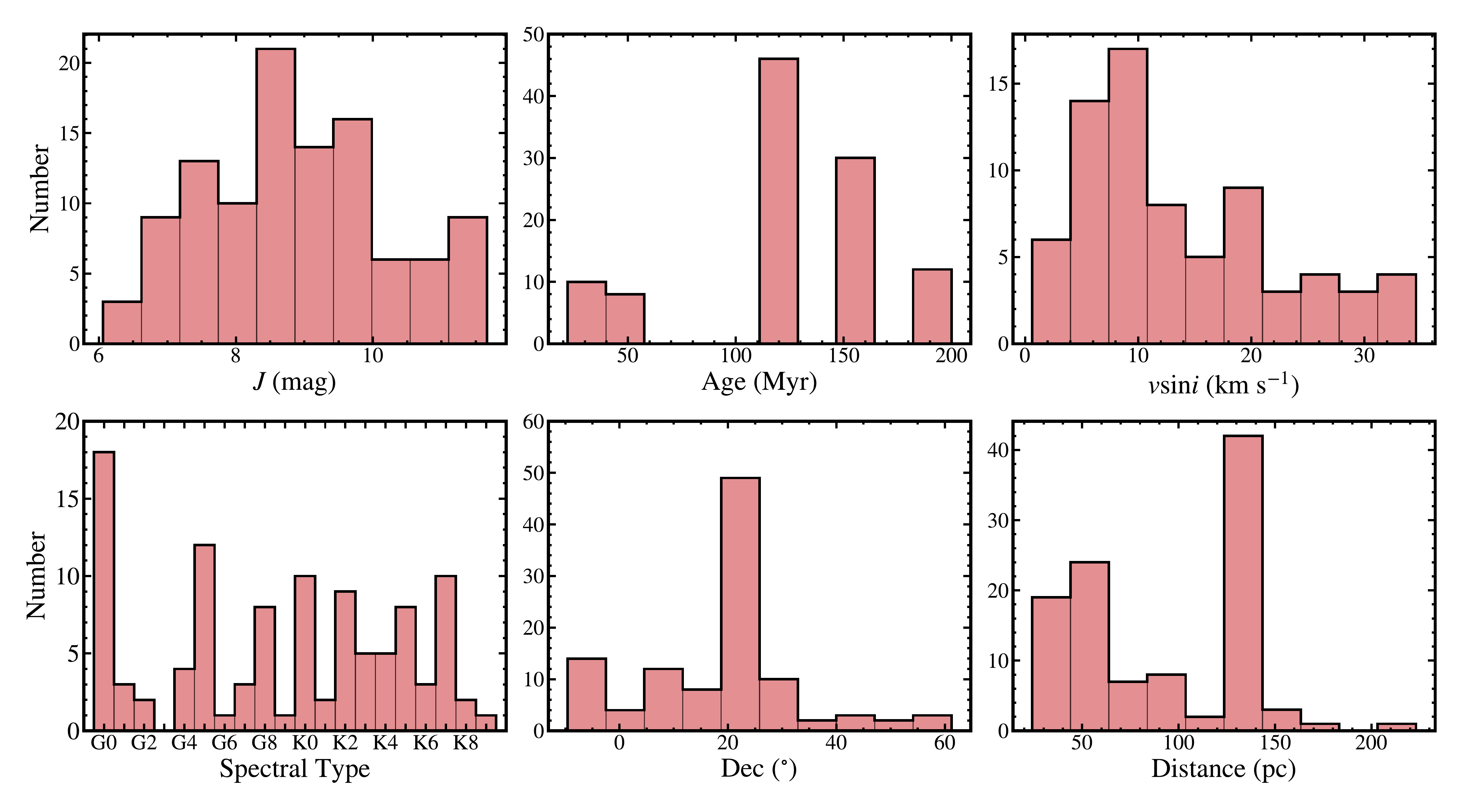}}
	    \caption{Characteristics of our full sample. The overdensities seen in the age, declination, and parallax distributions are from targets in the Pleiades, located at $\delta \approx +24^{\circ}$ and $\pi \approx 7$ mas with a common age of 120 Myr. Several targets have projected rotational velocities slightly beyond our 30 km s$^{-1}$ selection cut. These $v$sin$i$ measurements in the lower left panel are from our homogeneous recharacterization of our targets described in Section \ref{sec:vsini} with our HPF science observations.}
	    \label{fig:dist_summary}
    \end{figure*}

    Our final sample contains approximately 100 single young G and K dwarfs. The distribution of $J$--band magnitudes, spectral types, ages, declinations, projected rotational velocities, and parallaxes are displayed in Figure \ref{fig:dist_summary}. Our sample spans G0 to K9 spectral types, $J$--band magnitudes from 6.0 to 11.5 mag, and a stellar mass range of 0.7--1.2 M$_\odot$. The rotational velocity distribution is measured following the procedure described in Section \ref{sec:vsini}. Our $v$sin$i$ measurements for three stars slightly exceed our original threshold of 30 km s$^{-1}$ based on values from the literature, but all targets in our final sample ultimately have projected rotational velocities $<$35 km s$^{-1}$. In the age, declination, and parallax distributions, a large overdensity is apparent at $+24^{\circ}$, 140 pc, and 120 Myr, respectively, which arises from the large number of Pleiades members in our sample.
    
\begin{figure}[!ht]
	    \centerline{\includegraphics[width=1.0\linewidth]{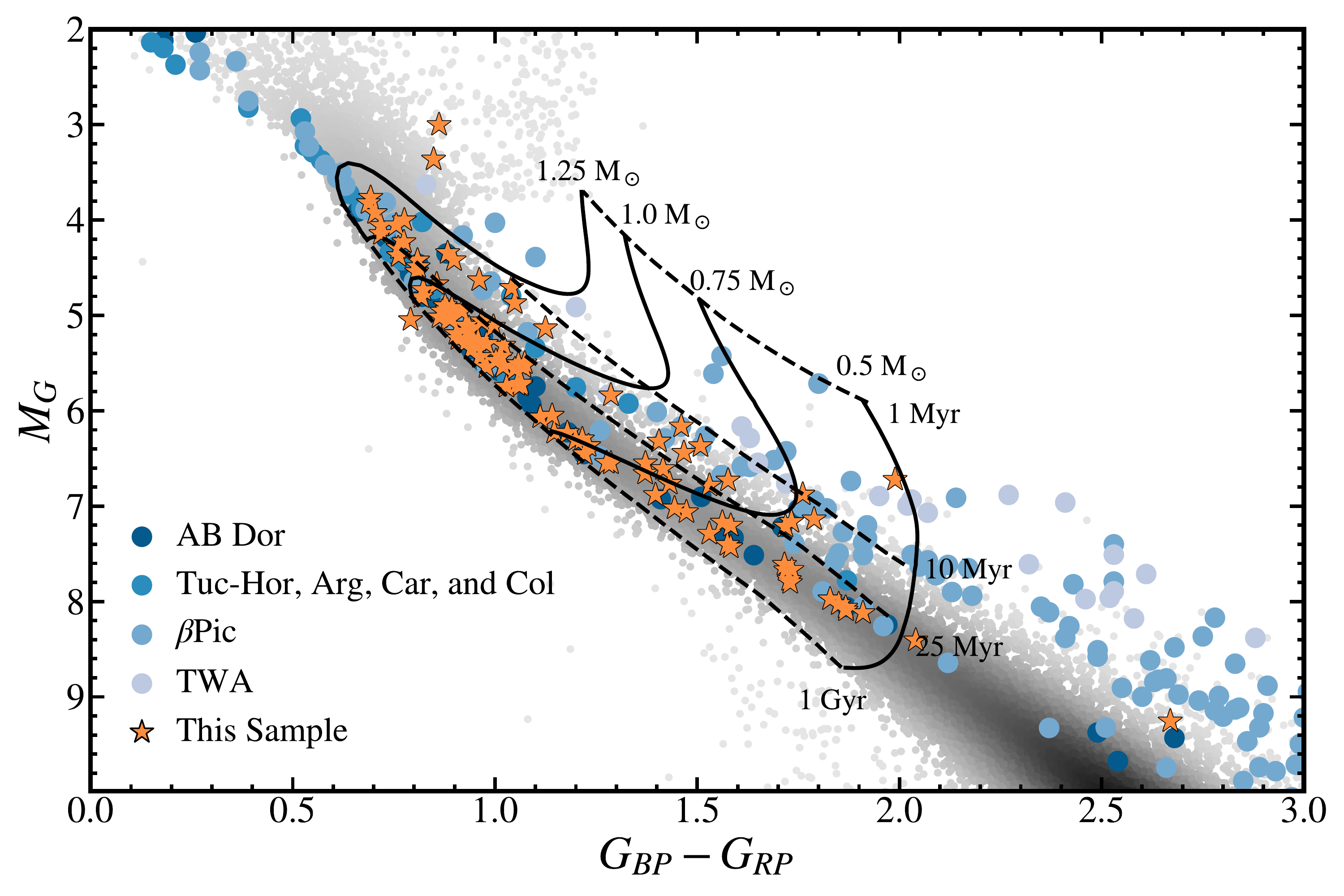}}
	    \caption{\textit{Gaia} DR2 color-magnitude diagram of stars within 100 pc (black dots) and our sample (yellow stars), compared to known moving group members from \citet{Malo2013} (blue circles). MIST isochrones at 1 Myr, 10 Myr, 25 Myr, and 1 Gyr are overlaid in solid black lines. Iso-masses in the range $0.5-1.25$ M$_\odot$ with steps of 0.25 M$_\odot$ are displayed as black dotted lines. All of our stars lie on or slightly above the main sequence, but individual ages are adopted from YMG membership studies. The 100 pc comparison sample is selected following the description in \citet{Bowler2019}.}
	    \label{fig:gaia_cmd}
    \end{figure}

    We also plot the \textit{Gaia} DR2 absolute $M_G$ versus $G_{BP} - G_{RP}$ color-magnitude diagram of our sample against stars within the nearest 100 pc in Figure \ref{fig:gaia_cmd}. We include 1 Myr--1 Gyr isochrones in the \textit{Gaia} bandpasses and 0.5--1.25 M$_\odot$ iso-masses from the MIST evolutionary tracks \citep{Paxton2011, Evans2018, Gaia2018}.
    
    This paper focuses on the first 14 months of our survey to assess the near-infrared stellar jitter of our sample of intermediate-age Sun-like stars, which will be refined as more data are obtained. To quantify these results, we model the underlying distribution of stellar jitter based on 29 stars from our sample with at least three epochs of RVs. Each epoch is made up of 3 contiguous measurements, which are combined using a weighted mean and the associated uncertainty.
    
\subsection{Observations} \label{sec:Data}

\begin{figure}[!htp]
	    \centerline{\includegraphics[width=1.00\linewidth]{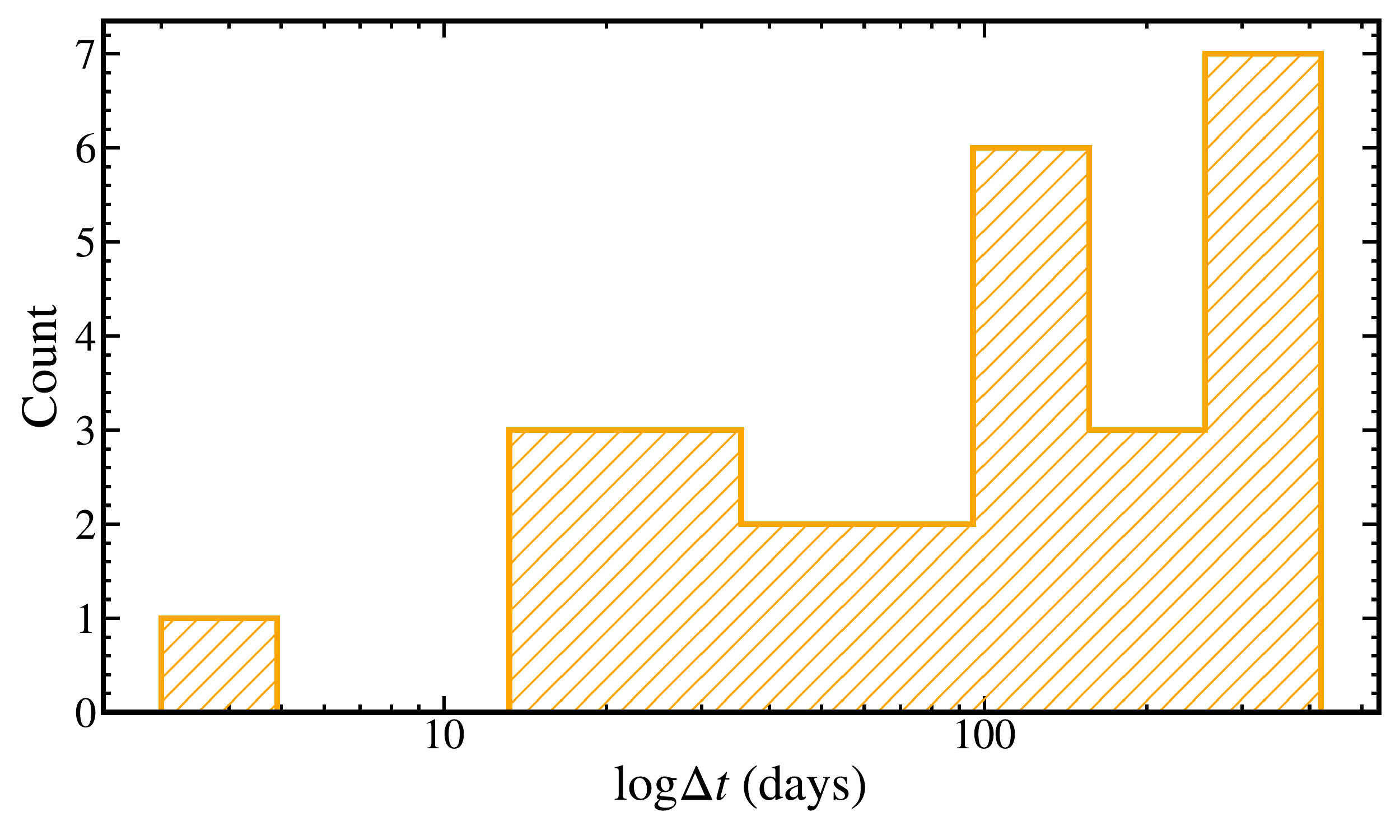}}
	    \caption{The distribution of time baselines for the subsample of 29 stars. All stars but one have a baseline more than 6 days, the typical expected rotation period of stars in our sample \citep{Mamajek2008}.}
	    \label{fig:time_base}
    \end{figure}

    Our HPF observations were taken in queue mode on a flexible scheduling system \citep{Shetrone2007}. Observations for individual targets are designed to sample both short and long intervals with semi-random cadence on timescales of days to years. For each RV epoch, 3 contiguous exposures are taken. Each exposure has a typical integration time of 300 seconds to maximize the S/N without saturating HPF's LFC (which has a maximum exposure time of 315 seconds). Typical visits therefore span approximately 15 minutes of consecutive integration. This observing strategy allows us to average over short-term stellar variations such as p-mode oscillations which peak at comparable frequencies \citep{Duvall1988, Anderson1990, Toutain1992, Appourchaux2008}. 29 targets with at least three epochs form the basis of the NIR activity measurement we focus on in this paper (see Section \ref{sec:nir_jitter}). $\sim$70\% of this sample have 3 epochs of RV measurements, with the rest having between 4--8 epochs. The distribution of time baselines for the survey subsample is shown in Figure \ref{fig:time_base}. Among 29 stars in our subsample, 28 have been observed over a time baseline that is longer than the typical expected rotation period ($P < 6$ days) for young, Solar-type stars \citep{Mamajek2008}.
    
    1D spectra are optimally extracted from the 2D spectral traces with the custom HPF data-extraction pipeline following the procedures outlined in \citet{Ninan2018}, \citet{Kaplan2018}, and \citet{Metcalf2019a}. Wavelength calibration is achieved using a optical frequency comb, which produces a regularly spaced array of laser pulses from 0.7--1.6 $\mu$m \citep{Metcalf2019b}. The HPF LFC is capable of correcting RV drifts for on-sky HPF observations at the $\sim$30 cm s$^{-1}$ level \citep{Stefansson2020}. HPF has a total of 28 echelle orders spanning 0.81--1.28 $\mu$m (\textit{z}, \textit{Y} and \textit{J} bandpasses). The orders are indexed 0--27, with the bluest order at 0 and the reddest order at 27. Throughout this study, wavelength ranges for these orders are described in air wavelengths. The orders have an average resolving power of $R \equiv \lambda$/$\Delta \lambda$ = 55,000 \citep{Mahadevan2012, Ninan2019}. Each order contains three optical fibers which simultaneously disperse the sky, the science target, and the laser frequency comb.
    
\section{\textbf{Precision Radial Velocity Pipeline}} \label{sec:Pipeline}

    We have developed a custom HPF precision RV pipeline designed to extract RVs of young G and K dwarfs for the Epoch of Giant Planet Migration Planet Search Program. \citet{Metcalf2019a} described one approach to extract precise RVs for M dwarfs with HPF, which follows the least-square matching (LSM) technique detailed in \cite{Guillem2012} and is based on the SpEctrum Radial Velocity Analyzer (\texttt{SERVAL}) code by \cite{Zechmeister2018}. However, the limiting RV precision for G and K dwarfs with HPF has not been thoroughly explored. For this reason, we investigated two different methods to extract RVs with our pipeline: LSM and a cross-correlation function (CCF). We found that the LSM approach ultimately yielded more precise RVs. Below we describe our implementation of the LSM technique and summarize the resulting RV performance for our survey.
    
    Prior to measuring RVs, all optimally-extracted 1D spectra are first corrected for the Earth's barycentric motion. This velocity shift is calculated using the \texttt{barycorrpy} \citep{Kanodia2018} package, a Python implementation of the barycentric correction algorithm by \citet{Wright2014}. This package applies corrections to better than 1 cm s$^{-1}$, as demonstrated by their comparison with the pulsar timing software \texttt{TEMPO2} \citep{Hobbs2006, Edwards2006}. \texttt{barycorpy} takes as input each target's proper motion, parallax, sky coordinates, and the observatory's position to account for the various factors influencing the barycentric motion (including secular acceleration; light-travel time; Earth's diurnal rotation, axial precession, and nutation; general relativistic effects from the Shapiro delay and gravitational blueshift from the Sun; gravitational redshifts from other solar system objects; and second order corrections to the non-relativistic Doppler formula).
    
    This correction is optimally applied to individual spectral orders. HPF's H2RG infrared array non-destructively reads up-the-ramp pixel intensity values during individual exposures to create 2D slope images which are extracted as 1D spectra. This nondestructive reading allows the instrument to calculate a true flux-weighted exposure time midpoint for each spectral order. These flux-weighted exposure time midpoints are further corrected for Roemer delay \citep{Eastman2010}. Barycentric correction is then applied to each order using the corrected exposure time midpoint. These corrections have been made for all spectra referred to in the proceeding sections, which are now in each target's reference frame.
    
\subsection{Template Creation for Least-Squares Matching}\label{sec:Template}
    
\begin{figure*}[ht]
        \vskip -0.04 in
	    \centerline{\includegraphics[width=1.0\linewidth]{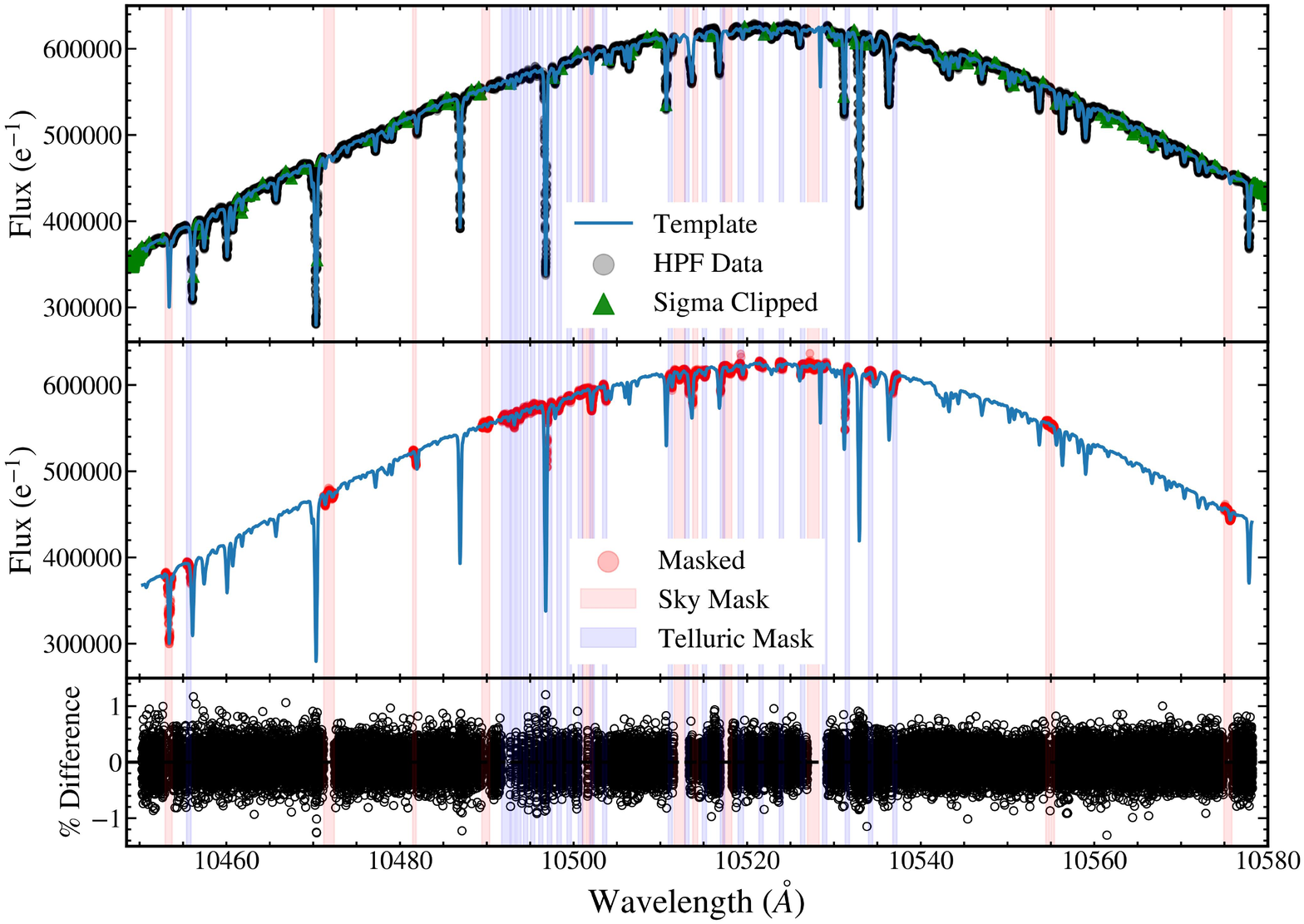}}
	    \caption{Example of the template creation for the RV standard HD 3765 over echelle Order 17. In the top panel, the HPF data are plotted in black circles and 3$\sigma$-clipped data are plotted in green triangles. In the middle panel, all data that were masked are plotted in red circles. The bottom panel shows the ratio of the science spectrum divided by the template. In the bottom panel, the RMS of the fractional difference between the template and the data is 0.26\%. In all three panels, the regions where the sky and telluric masks are applied are shaded in red and blue, respectively.}
	    \label{fig:hd_3765_o17}
    \end{figure*}

\begin{figure*}[ht]
        \vskip -0.04 in
	    \centerline{\includegraphics[width=1.0\linewidth]{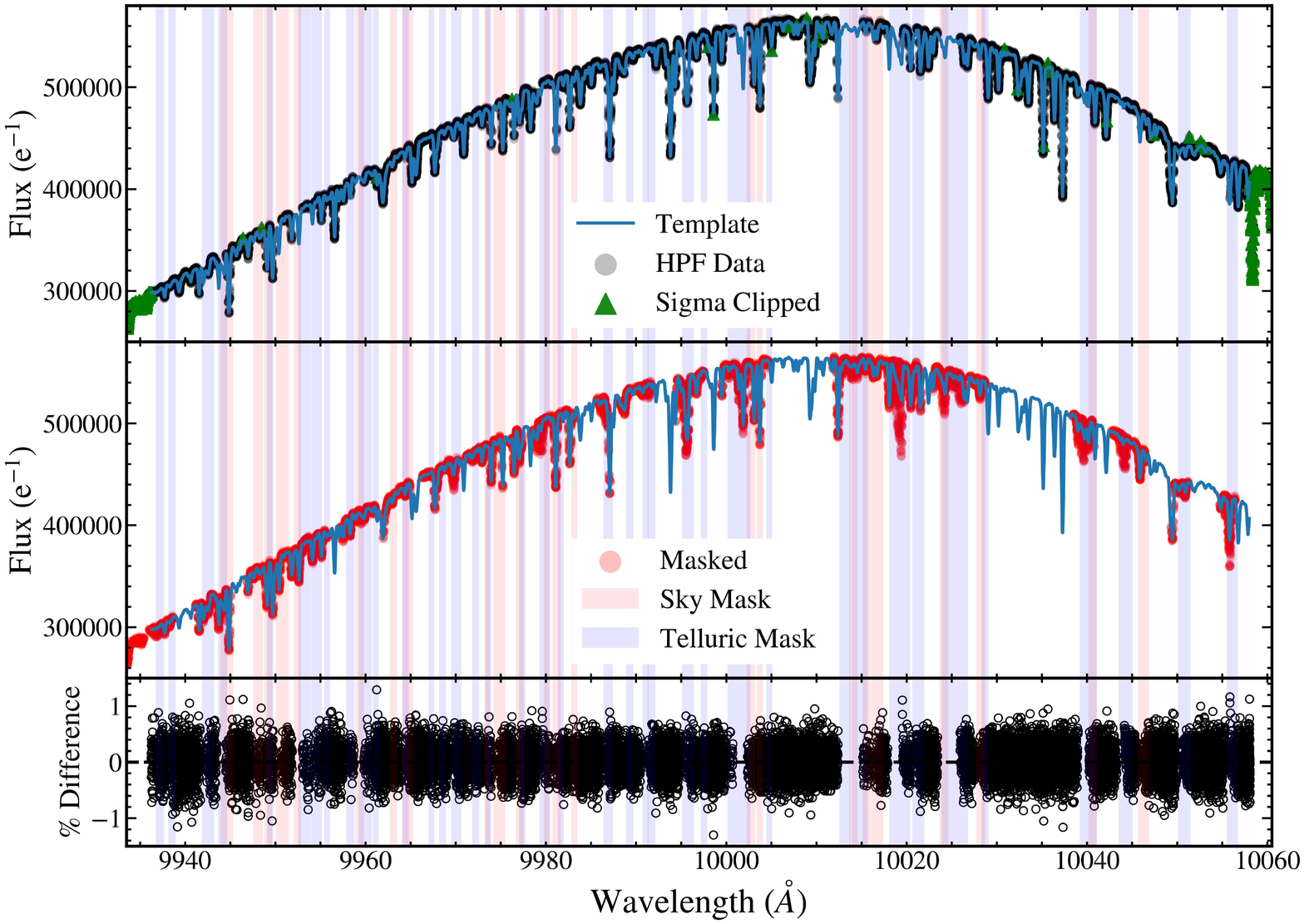}}
	    \caption{Example of the template creation for the RV standard HD 3765 over echelle Order 14. The panels are the same as in Figure \ref{fig:hd_3765_o17}. The RMS of the fractional difference between the template and the data in the lower panel is 0.27\%. Note that this order has many more telluric and sky lines than Order 17 shown in Figure \ref{fig:hd_3765_o17}.}
	    \label{fig:hd_3765_o14}
    \end{figure*}

    The framework for our LSM RV reduction pipeline is adapted from the publicly available \texttt{SERVAL} code to fit our needs \citep{Guillem2012, Zechmeister2018}. First, high frequency pixel-to-pixel variations are removed using a deblazed master flat spectrum. Then, a median coadded night sky spectrum taken on the same night is subtracted. An initial template is created by multiplicatively adjusting all spectra to the same level as the highest S/N spectrum. Multiplicative 3rd-order polynomials which scale each spectrum on a pixel-to-pixel basis to this level are applied to account for any flux variation between observations. Once all spectra are normalized to a common scale, a uniform cubic basic (\textit{B}-spline) regression is applied on all spectral data points to create a master template. The best coefficients for the spline regression are computed by minimizing the residuals between the template and science spectra using the following modified weighted $\chi^2$ value:
    \begin{equation} \label{eq:temp_chi2}
        \chi^2 = \sum_n\sum_{i} \frac{p_{n,i}^2}{\sigma_{n,i}^2} \left[\frac{S_{n,i}}{p_{n,i}} - T(b) \right]^2,
    \end{equation}
    where $p_{n,i}$ is the polynomial for each observation $n$ at each pixel $i$, $S_{n,i}$ is the flux of the science spectrum $n$ at pixel $i$, $\sigma_{n,i}$ is the associated flux error, and $T(b)$ is the template regression described by the spline coefficient $b$. The template regression is oversampled and the template spline has $K$ knots, with the optimal knot spacing and number assessed using the Bayesian Information Criterion (BIC). The template undergoes two iterations of $\kappa$-sigma-clipping in which linear residuals between the data and the template exceeding a threshold ($r_i$) are removed, where
    \begin{equation}\label{eq:sig_clip}
        r_i > \kappa \epsilon_i \sqrt{\chi^2_\mathrm{red}}.
    \end{equation}
    Here $\kappa$ is a coefficient of value 3, $\epsilon_i$ is the error for pixel $i$, and $\chi^2_\mathrm{red}$ is the reduced chi squared statistic for the template minimization. 150 data points near the edge of the science spectrum are also removed to account for any edge effects. Using the optimal template constructed with the best fit $b$ parameter found by minimizing Equation \ref{eq:temp_chi2}, the best coefficients of the background polynomials and the RV shifts for each spectrum are then found simultaneously in the same error-weighted linear least-squares matching (LSM) fashion. Examples of this template creation process over two echelle orders (17 and 14) for the RV standard HD 3765 are shown in Figures \ref{fig:hd_3765_o17} and \ref{fig:hd_3765_o14}. Order 17 (10447--10581 \AA) is one of the least contaminated orders used, while Order 14 (9933--10061 \AA) is one of the most contaminated.
    
    Following \citet{Zechmeister2018}, the template is Doppler shifted over a range of velocities $v_k$; the residual at each pixel $i$ at each shift $k$ between the template and each observation $n$ is computed as follows:
    \begin{equation}
        \chi_k^2 = \sum_{i} \frac{\left[S_i - p_{i}(a) \cdot T_i(\lambda_k, b) \right]^2}{\sigma_{i}^2},
    \end{equation}
    where $a$ are the polynomial coefficients and $\lambda_k$ is the 3D Doppler shift formula \citep{Wright2014}, 
    \begin{equation}\label{eq:doppler}
        \lambda_k = \lambda\left({\frac{1 - \frac{v_b}{c}}{1 + \frac{v_k}{c}}}\right),
    \end{equation}
    where $v_b$ is the barycentric velocity, as calculated in Section \ref{sec:Pipeline}. All polynomial coefficients and velocity shifts are iteratively adjusted until a global minimum is reached. Velocities are shifted by $\Delta v$ steps of 50 m s$^{-1}$ from $-5$ km s$^{-1}$ to 5 km s$^{-1}$. The RV, $v$, for a spectrum is taken to be the minimum of the $\chi^{2}-$velocity curve, $v_m$, located at index $m$. As the template is cubicly interpolated, both the template itself and the $\chi^{2}-$velocity curve have continuous second derivatives. Following \citet{Zechmeister2018}, we leverage this to improve the estimate of the velocity minimum and its error by interpolating the $\chi^{2}-$velocity parabola. A better estimate of the RV is then
    \begin{equation}
        v = v_m - \frac{\Delta v}{2} \cdot \frac{\chi^2_{m+1} - \chi^2_{m-1}}{\chi^2_{m-1} - 2\chi^2_m + \chi^2_{m+1}},
    \end{equation}
    with a variance of
    \begin{equation}
        \epsilon_v^2 = \frac{2}{\left( \chi^2 \right)^{\prime\prime}} = \frac{2\Delta v^2}{\chi^2_{m-1} - 2\chi^2_m + \chi^2_{m+1}}.
    \end{equation}
    This process is performed on each suitable spectral order. The RV for each epoch is determined from the weighted mean of the spectral order RVs and the uncertainty is the standard error of these values.

    For the RV extraction process, only 8 echelle orders out of 28 are used due to the presence of strong and prevalent telluric absorption in other orders, making them impractical for measuring high precision RVs. We retain Orders 4 (8535--8645 \AA), 5 (8656--8768 \AA), 6 (8781--8895 \AA), 14 (9933--10061 \AA), 15 (10099--10229 \AA), 16 (10270--10402 \AA), 17 (10447--10581 \AA), and 18 (10630--10767 \AA), with indexing beginning at 0. These correspond to wavelength ranges in $z$ band (8535--8895 \AA) and $Y$ band (9933--10767 \AA).
    
    All spectral regions affected by telluric absorption lines and sky emission lines are treated following the strategy used in \citet{Zechmeister2018}. During the master template creation, they are down-weighted by a large factor of 10. Without the contribution from stellar activity, the stellar spectrum is invariant in the star's reference frame, but telluric absorption and sky emission lines caused by Earth's atmosphere will shift seasonally and from observation to observation as changing atmospheric conditions alter the strength of the telluric absorption and OH emission. Thus, in the template creation process, if telluric and sky emission lines are masked then large gaps in the template spectrum will appear as the lines shift over time due to the barycentric velocity variations over a season. Within these gaps no constraints are placed on the template and this lack of information can lead to substantial uncertainty in the creation of the template because the regression blindly attempts to fit this region. As more data are obtained over different seasons, these gaps are slowly filled and the template progressively improves in those regions \citep[see, e.g.,][]{Bedell2019}. This leads to better RV precision over time for all epochs of that particular target. Note that when an RV is measured for each epoch, the telluric and sky lines---which are located at unique velocity shifts corresponding to the date and time of the observation---are explicitly masked.
    
    The binary telluric, sky emission, and stellar outlier masks we use in our pipeline are produced by the HPF instrument builders as part of the spectral extraction as discussed in \citet{Metcalf2019a} and \citet{Stefansson2020}. The telluric mask is created using the publicly available package \texttt{TelFit} \citep{Gullikson2014}, a Python wrapper to the Line-By-Line Radiative Transfer Model (\texttt{LBLRTM}) package \citep{Clough2014}. A synthetic spectrum over the HPF spectral range is created using the default TelFit parameters of 50\% humidity at 30.6 \degree N and 2100 m altitude for the HET. Every region in the model with a transmission below 99.5\% is treated as a telluric in the binary mask. The binary mask is broadened by 17 pixels (0.459 \AA) to avoid smoothing the synthetic spectrum to HPF's resolution in order to account for shallower telluric lines. A sky emission line binary mask is created using a master 10 minute sky-background exposure. The background continuum is subtracted from this deep sky exposure and all regions above 5$\sigma$ are treated as sky emission lines.
    
\begin{figure}[!htp]
	    \centerline{\includegraphics[width=1.00\linewidth]{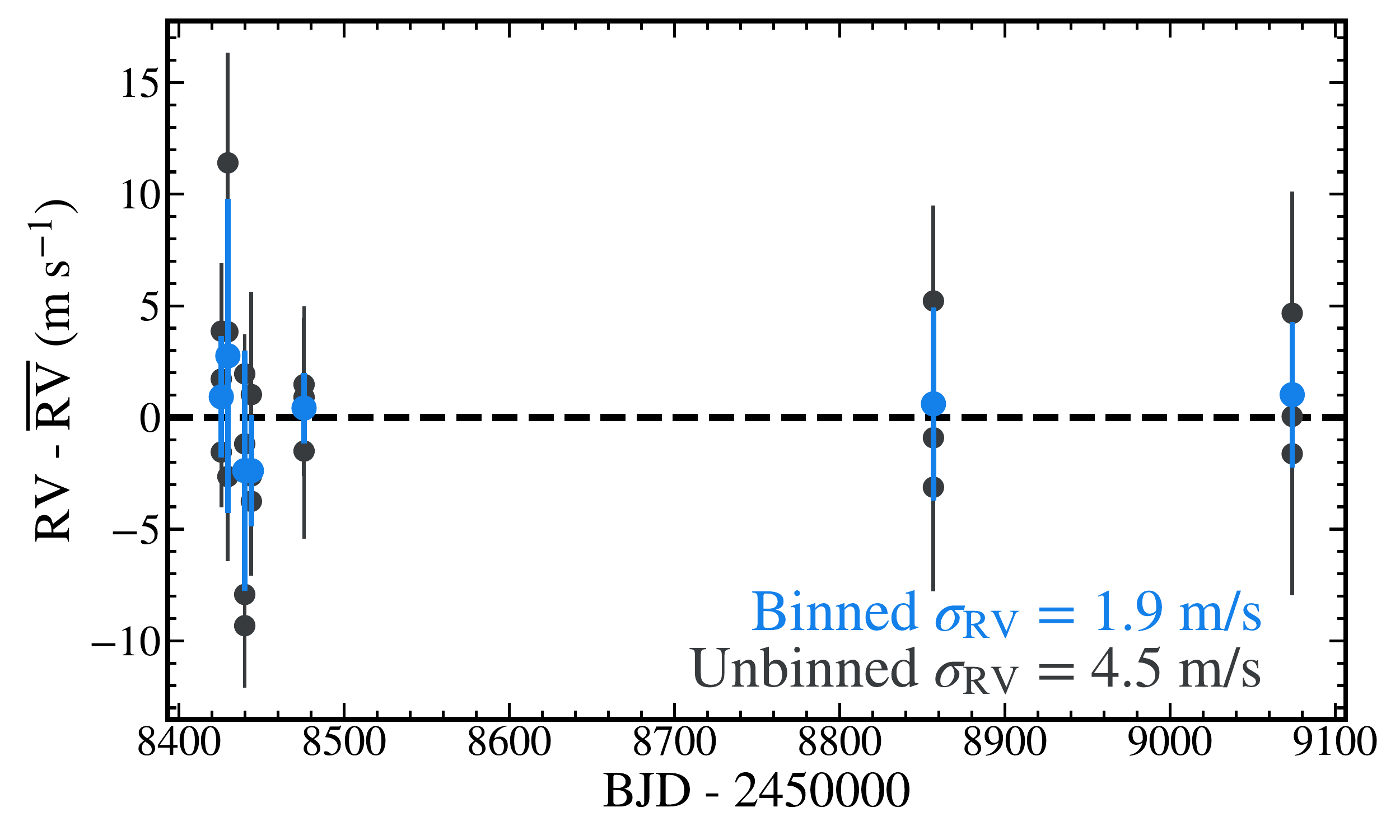}}
	    \caption{Our HPF RVs for the RV-stable standard HD 3765. Black points show RVs for the three contiguous spectra we obtain at each epoch. These points are combined into a weighted average, shown in blue. We measure a binned RV RMS precision of 1.9 m s$^{-1}$, demonstrating both instrumental stability and pipeline precision. This precision is comparable to the RMS value of 2.4 m s$^{-1}$ in the optical for the same target \citep{Isaacson2010}.}
	    \label{fig:hd_3765}
    \end{figure}

    Using this technique, we measure 1.9 m s$^{-1}$ (RMS) binned precision on the bright ($J$=5.7 mag) K2 RV standard star HD 3765  (Figure \ref{fig:hd_3765}), which is slightly better than the 2.4 m s$^{-1}$ value from optical measurements for the same target \citep{Isaacson2010}.
    
    We compare the RV precision from the LSM technique and a CCF approach using synthetic spectra from the PHOENIX model grids \citep{Husser2013} and ultimately adopt the LSM method for measuring precise RVs in this survey. Using the CCF approach on the same RV standard, HD 3765, we measure a nightly binned precision of 2.9 m s$^{-1}$. This is 53\% larger than our measurement using the LSM technique. This is consistent with tests using RV pipelines for other spectrographs and likely originates from the fact that the CCF does not extract all RV information in an optimal way \citep[e.g.,][]{Chelli2000, Bouchy2001, Pepe2002, Guillem2012}. The CCF technique assumes the model is a perfect representation of the data and does not fully capture asymmetries in the line profile. The precision and accuracy of the CCF method depends on the model used, which is often simply a binary mask based on carefully chosen spectral lines. As the LSM technique works with the entire spectrum, it utilizes all available RV information. For later-type stars, this is especially advantageous as there is a wealth of RV information due to an increased number of stellar lines \citep{Guillem2012}. In our own measurements, we notice this limit in precision and have chosen the LSM technique as the preferred method of RV extraction.

\subsection{Stellar Characterization and Rotational Velocities}\label{sec:vsini}

    In addition to extracting RVs, we use the science spectra to homogeneously characterize each target. In particular, we focus on determining the projected rotational velocity ($v$sin$i$) of every star, which is expected to correlate with limiting per-epoch RV precision and the activity levels of our targets.
    
    All science spectra are first normalized through the following procedure. High frequency variations are first removed from the science spectrum using a deblazed master flat spectrum created by averaging multiple flats taken during a single night. Then, the blaze function is removed using a median co-added master flat image created from the flat field images of the corresponding night. This is done by scaling each flat spectrum to the same scale as the median flat image by a multiplicative factor, determined using the formalism described in \citet{Cushing2008},
    
    \begin{equation}\label{eq:cushing_median}
        C = \frac{\sum_i F_i M_i/\sigma_i^2}{\sum_i M_i^2/\sigma_i^2}.
    \end{equation} \\
    Here, $i$ refers to the $i$th pixel, $M_i$ is the flux for the median flat spectra, and $F_i$ is the flux for the flat spectrum. The median flux value of all scaled spectra is taken at each pixel. The continuum is then defined by binning the spectra and taking the median of the highest flux values in order to eliminate the contribution of absorption lines. This continuum level is then used to normalize the spectra with a 2\textsuperscript{nd}-order Legendre polynomial fit. This process is repeated for each of the eight HPF orders used in our RV measurement.
    
    Spectroscopic parameters ($T_\mathrm{eff}$, log$g$, [Fe/H]) and rotational velocities ($v$sin$i$) are derived using these normalized orders through the publicly available package \texttt{iSpec} \citep{Blanco-Cuaresma2014, Blanco-Cuaresma2019}. All normalized spectra are corrected for the barycentric motion of the Earth, as calculated in Section \ref{sec:Pipeline}. They are further corrected for the absolute RV using a simple CCF match to a synthetic solar template.
    
    A synthetic stellar spectrum is created over the wavelength range of all orders used for RV extraction. Instead of using the standard line list recommended by \texttt{iSpec}, we adopt the atomic line list from the Vienna Atomic Line Database (VALD) \citep{Kupka2011}, which spans a broader wavelength range of 3000 to 11000 \AA. The MARCS models are used as the model atmosphere grids \citep{Gustafsson2008} and solar abundances are taken from \citet{Grevesse2007} in order to match the MARCS grids. The stellar spectral synthesis is handled by the open source radiative transfer code \texttt{SPECTRUM} \citep{Gray1994}.
    
    \begin{figure*}[tb]
	    \centerline{\includegraphics[width=1.025\linewidth]{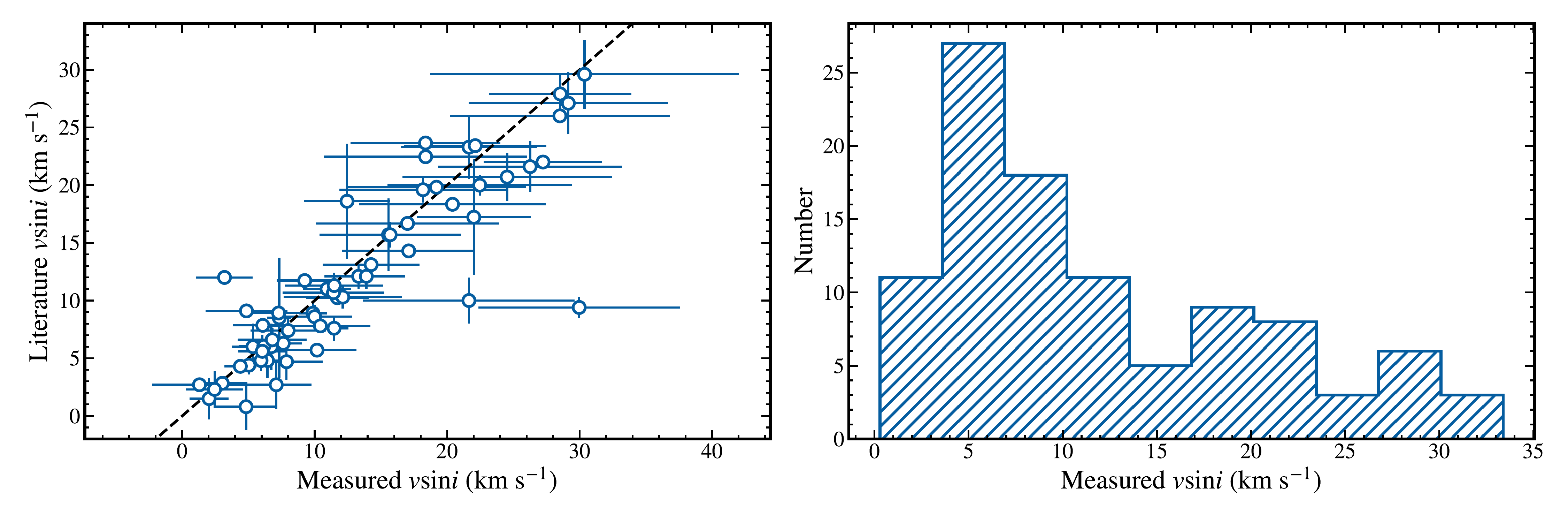}}
	    \caption{\textbf{Left:} Comparison between our $v$sin$i$ measurements from HPF spectra following the procedure described in Section \ref{sec:vsini} with values for the same targets from the literature. Only the 75 stars that have literature values are plotted. The black dashed line shows the 1:1 relation. Our measurements are in good agreement with literature values. \textbf{Right:} The distribution of measured rotational velocities ($v$sin$i$). Our targets span $v$sin$i$ values from 1.0 to 34.6 km $^{-1}$, with 78\% being between 5--20 km $^{-1}$.}
	    \label{fig:vsini}
    \end{figure*}
    
    \texttt{iSpec}'s synthetic spectral fitting technique minimizes the $\chi^2$ value between the normalized observed spectra and the synthesized model spectra. Initial atmosphere parameters are taken from the literature. If there are no previous atmospheric measurements, then solar parameters are assumed. We set the microturbulence to $V_\mathrm{micro}=1.15$ km s$^{-1}$ and the macroturbulence to $V_\mathrm{macro} = 2 V_\mathrm{micro}$, which is based on values typical of young ($\lesssim 250$ Myr) F, G, and K stars \citep{Folsom2016, Baratella2020}. Varying these values does not substantially change the results of the spectral fit. The only free parameters in the fit are the atmospheric parameters and rotational velocity. Everything else in the model, including the chemical abundances, statistical weights, and oscillators strengths are kept static. Figure \ref{fig:vsini} shows $v$sin$i$ values from the literature compared to our measured values from this procedure for the 75 targets in our sample with literature values. There is excellent agreement between literature and measurement values.
\section{\textbf{Stellar Activity at Intermediate Ages}} \label{sec:Jitter}
    Observing at near-infrared wavelengths is expected to reduce the RV contributions of starspots because spots are cooler than the surrounding photosphere by hundreds to thousands of Kelvin. The difference in resulting contrast between the photosphere and the starspot is greater at optical wavelengths than in the near-infrared. As stellar rotation moves these spots in and out of view, this enhanced contrast results in larger RV variations at optical wavelengths \citep{Prato2008, Mahmud2011, Crockett2012}. In this study we measure NIR jitter\footnote{We broadly apply the term jitter to the combined effects of the RV variability due to intrinsic stellar photospheric motion, its effect on emergent stellar spectra, and both instrumental and photon-limited observational uncertainties.} for a significant sample of Sun-like stars at intermediate ages for the first time. We compare our findings with those of similar surveys in the optical below.

\subsection{Stellar Jitter in the Optical for Young Stars}
\label{sec:optical_jitter}

    \citet{Hillenbrand2015} measured the stellar jitter for a subset of 171 stars from the California Planet Search program \citep{Wright2012}. Their sample selected for nearby stars with Sun-like masses (0.8--1.2 $M_\odot$) and ages younger than the Sun. They measured the chromospheric activity level index, $R'_{HK}$ \citep{Noyes1984}, for each of their targets. Using the median activity-age relationship from \citet{Mamajek2008}, they convert activity to ages and demonstrate that stellar jitter scales approximately logarithmically with age. At the typical ages of our sample ($\sim$20--200 Myr), the corresponding intrinsic stellar activity level in the optical is expected to be $\sim$60 m s$^{-1}$.
    
    Inspired by this work, we assemble a sample of optical RV measurements of stars with properties consistent with those in our survey in order to directly compare optical and NIR jitter. One of the earliest precision RV surveys of young stars was carried out by \citet{Paulson2006}, who targeted 61 G, K, and M stars with ages ranging from $\sim$10 to $\sim$300 Myr in nearby young moving groups using the MIKE spectrograph on the Magellan Telescopes. Applying our stellar age, rotation rate, and spectral type selection cuts to their sample, 31 objects are comparable to our sample. They report an $A_{obs}$ value, which is the RMS of the RVs for each star and directly corresponds to the jitter that we report. We further remove one star, HIP 29964, from this sample as its $A_{obs}$ value was a 3$\sigma$ outlier compared to the rest of the population. The median and mode RV RMS of these final 30 targets is 60.0 and 58.0 m s$^{-1}$, respectively. 
    
    \citet{Lagrange2013} conducted a giant planet RV survey of young stars with HARPS. Their program targeted 26 A--K dwarfs with ages ranging from 8 to 300 Myr. Of these, 8 were G and K dwarfs with ages and rotational velocities matching our survey. The median and mode RV jitter of these stars is 63.2 and 69.0 m s$^{-1}$, respectively. More recently, \citet{Grandjean2020} presented results from their HARPS RV search for young planetary systems. Their survey is designed to measure the occurrence rate of giant planets with orbital periods up to 1000 days. Their stellar sample consists of 89 A--M type young ($<$300 Myr old) stars found within 80 pc. Within this sample, 20 targets are inclusive of our spectral, age, and $v$sin$i$ selection ranges. The median RV jitter of this sample is 56.5 m s$^{-1}$ and mode is 45.0 m s$^{-1}$.
    
    Since there is no overlap of targets among these three surveys, we combine them into a total sample of 59 optical jitter measurements. The average age of the sample is 109 Myr and the median and mode RV RMS is 60 m s$^{-1}$ and 55.7 m s$^{-1}$, respectively. The 68\% and 95\% credible intervals are 3--165 Myr and 3--187 Myr, respectively. The optical jitter at these ages is consistent with the relationship found by \citet{Hillenbrand2015}. We plot the combined optical RV jitter distribution in Figure \ref{fig:rv_jit_comp}.

\subsection{Stellar Jitter in the NIR for Young Stars}
\label{sec:nir_jitter}

    Depending on the spectrograph resolving power and wavelength grasp, the NIR RV precision can be lower than the corresponding optical RV precision due to the smaller number of spectral lines that can be used to extract RV information. Both \citet{Figueira2016} and \citet{Reiners2020} demonstrated that NIR precision does not substantially improve above 1 $\mu$m for M dwarfs. This survey focuses on G and K dwarfs, which have even fewer lines at those wavelengths. However, HPF's spectral range reaches down to 0.81 $\mu$m to address this issue. In the red optical $z$ band, G, K, and M dwarfs have a higher density of atomic and molecular absorption lines, greatly mitigating this issue.
    
\begin{figure}[!tp]
	    \centerline{\includegraphics[width=1.025\linewidth]{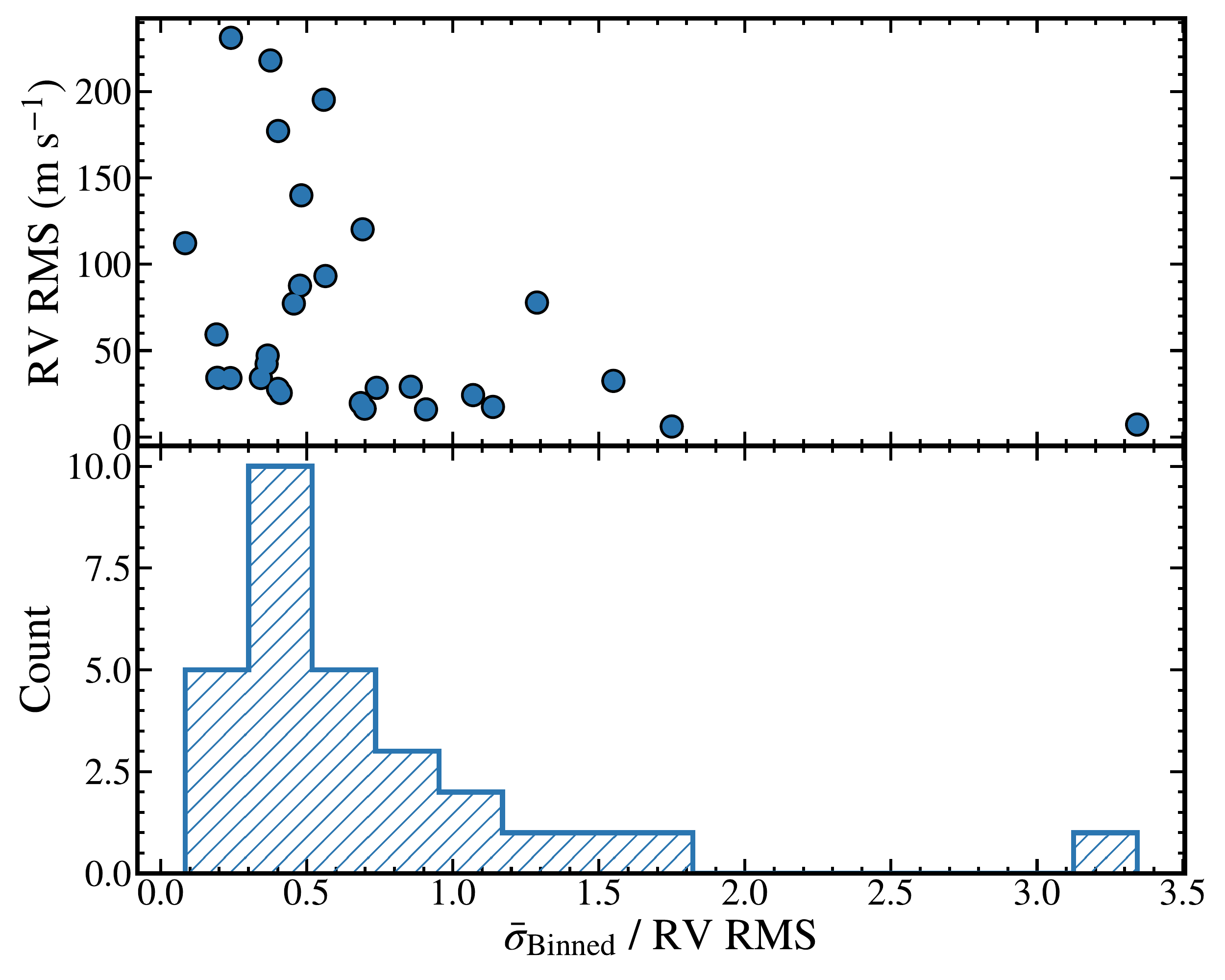}}
	    \caption{The ratio of average NIR RV measurement error ($\bar{\sigma}_\mathrm{Binned}$) to the RV RMS measurement for targets with $N \geq 3$ observations. $50\%$ of targets have an error-to-RMS ratio less than 0.5, with only six targets above a ratio of 1.0. These four stars all have RV RMS values below 25 m s$^{-1}$; we expect this to be limited by our measurement precision rather than jitter.}
	    \label{fig:err_avg_ratio}
    \end{figure}

    Among our full sample of targets, 29 stars have at least three epochs of randomly sampled RV measurements. Our per epoch RV precision is determined by the limiting instrument precision and the properties of the target, such as brightness, spectral type, and $v$sin$i$. We assess the overall level of NIR RV precision of this subsample by comparing the average measurement error ($\bar{\sigma}_\mathrm{Binned}$) to the measured RV RMS for each target. If the RV uncertainty is lower than the intrinsic RV jitter, then the ratio of the measured RV uncertainty to the measured RV RMS should be below 1. When the intrinsic jitter is unusually high, this ratio is well below 1. Our ability to determine stellar jitter will be limited by the RV measurement precision if this ratio is greater than 1. In our subsample, this only happens for targets where the RV RMS is small. We plot the ratio of RV error to RV RMS in Figure \ref{fig:err_avg_ratio}; 79\% of our sample has a $\bar{\sigma}_\mathrm{Binned}/\mathrm{RMS} < 1$, indicating that RV precision does not play a dominant role in the RV RMS measurement, and therefore RV RMS is a good indicator of the intrinsic stellar jitter.
    
\begin{figure}[!tp]
	    \centerline{\includegraphics[width=1.025\linewidth]{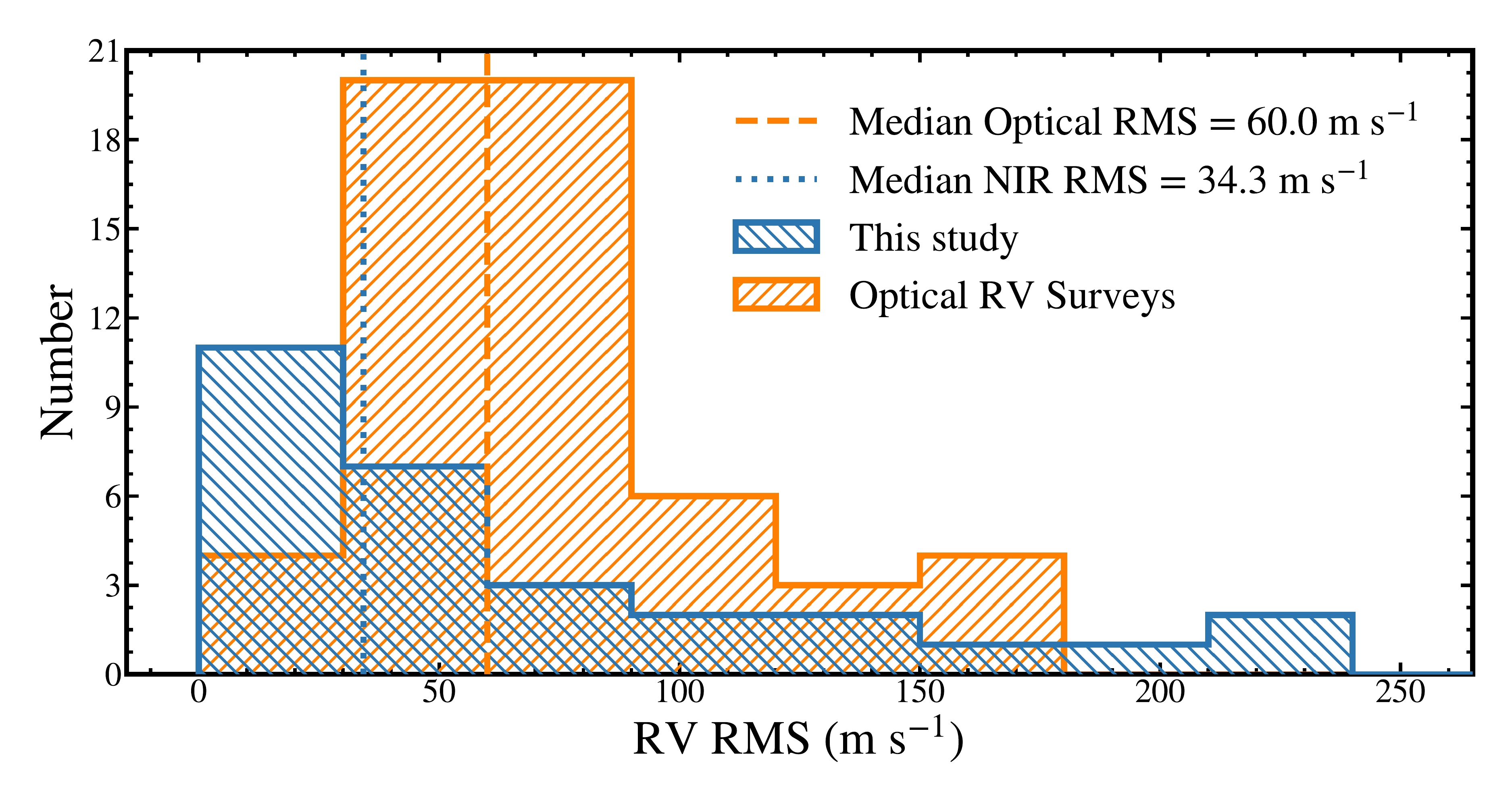}}
	    \caption{The optical RMS distribution of young Sun-like stars from literature (orange) compared with our measurements with HPF in this study (blue) over the same range of age (20--200 Myr), spectral type (G and K), and rotational velocity (1.0--35.0 km s$^{-1}$). The median NIR RV RMS (blue dotted line), 34.2 m s$^{-1}$, is reduced by a factor of $\sim$2 from the median optical RV RMS (orange dashed line) of 60.0 m s$^{-1}$. Note that the optical sample is approximately twice as large as the NIR sample, but this difference in sample size only influences the standard error of the median value as a sample statistic.}
	    \label{fig:rv_jit_comp}
    \end{figure}

\begin{figure*}[!tp]
	    \centerline{\includegraphics[width=1.025\linewidth]{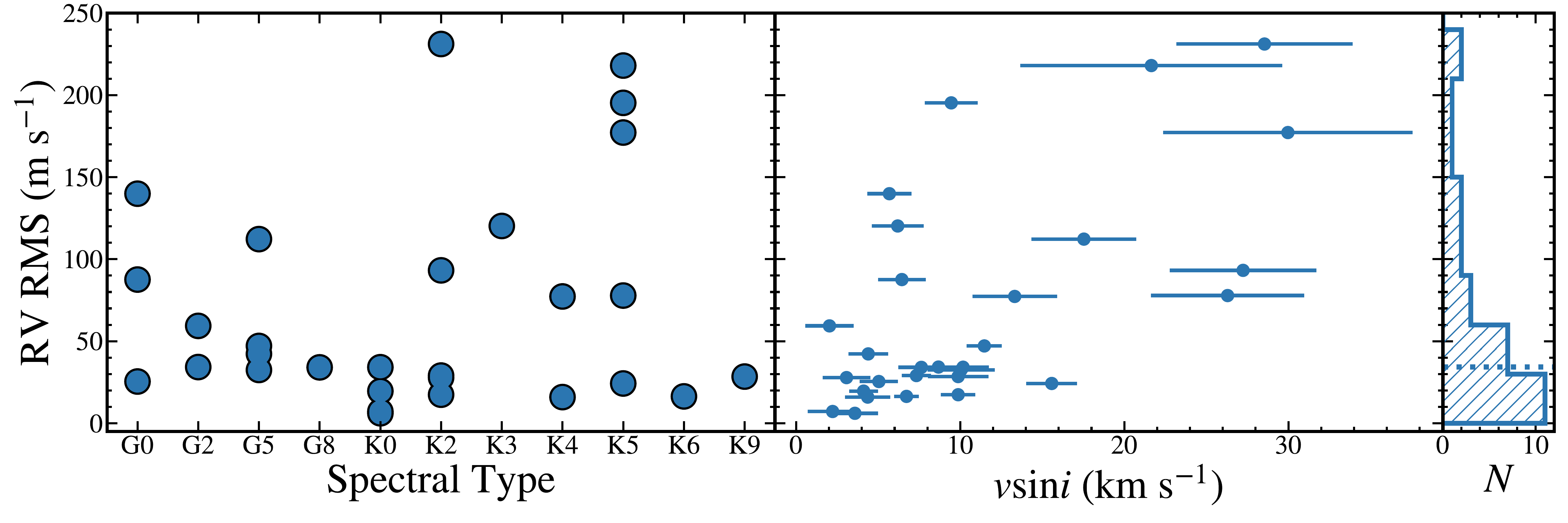}}
	    \caption{\textbf{Left:} NIR RV RMS distributions for targets with $N \geq 3$ observations from this study as a function of spectral type. \textbf{Center:} The same measurements as a function of $v$sin$i$. \textbf{Right:} The distribution of NIR RV RMS. The blue dotted line is the median NIR RV RMS of 34.2 m s$^{-1}$.}
	    \label{fig:rv_spt}
    \end{figure*}

    This subsample is also a large enough sample size to make initial inferences about the NIR RV jitter properties of young Sun-like stars from 20--200 Myr. We plot the distribution of NIR RV RMS for our intermediate-age stars in blue in Figure \ref{fig:rv_jit_comp} and as a function of stellar spectral type in Figure \ref{fig:rv_spt}. The median and mode RV RMS in the NIR is 34.2 m s$^{-1}$ and 32.3 m s$^{-1}$, respectively. The 68\% and 95\% credible intervals are 6.1--95.9 m s$^{-1}$ and 6.1--201.7 m s$^{-1}$, respectively. Table \ref{tab:sample_info} in Appendix \ref{sec:appendix} reports the properties and measured RV RMS of all 29 stars.
    
    Below we explore simple parametric models for the underlying distribution of jitter levels for our sample.
    

\subsection{Modelling the Stellar Jitter Distribution} \label{sec:stat_jitter}
    
    We use a Markov Chain Monte Carlo (MCMC) algorithm to fit our observed RV RMS distribution with an underlying model for the near-IR jitter. Although the functional form of the intrinsic distribution is unknown, using a constant value to describe the population of stellar jitter is unphysical because the distribution of ages, masses, and rotational velocities of stars in our sample should produce a spread in activity levels. Our approach is to model the underlying NIR jitter as a log-normal distribution to reflect the fact that it is bounded by zero on one side and right skewed. Moreover, log-normal distributions are often good models when values cannot be negative, mean values are low, and variances are large \citep[e.g.,][]{Limpert2001}. We choose this model for stellar jitter due to its flexibility, small number of shape parameters, and how well it reproduces both observed optical and NIR RV RMS distributions. The log-normal distribution is characterized by two parameters, the jitter mean ($\mu_\mathrm{jit}$) and the jitter standard deviation ($\sigma_\mathrm{jit}$):
    \begin{equation}
        P(\mathcal{S}) =  e^{\mathcal{N}(\mu_\mathrm{jit},\,\sigma_\mathrm{jit}^{2})}.
    \end{equation}
    Here $\mathcal{S}$ represents the spread in the jitter of each star, which we describe in more detail below. In this model, comparable values of $\sigma_\mathrm{jit}$ and $\mu_\mathrm{jit}$ will produce narrow distributions of jitter values, whereas large differences will produce a broad range of jitter levels.

\begin{figure}[!tp]
        \hspace{-0.3cm}
	    \centerline{\includegraphics[width=1.0\linewidth]{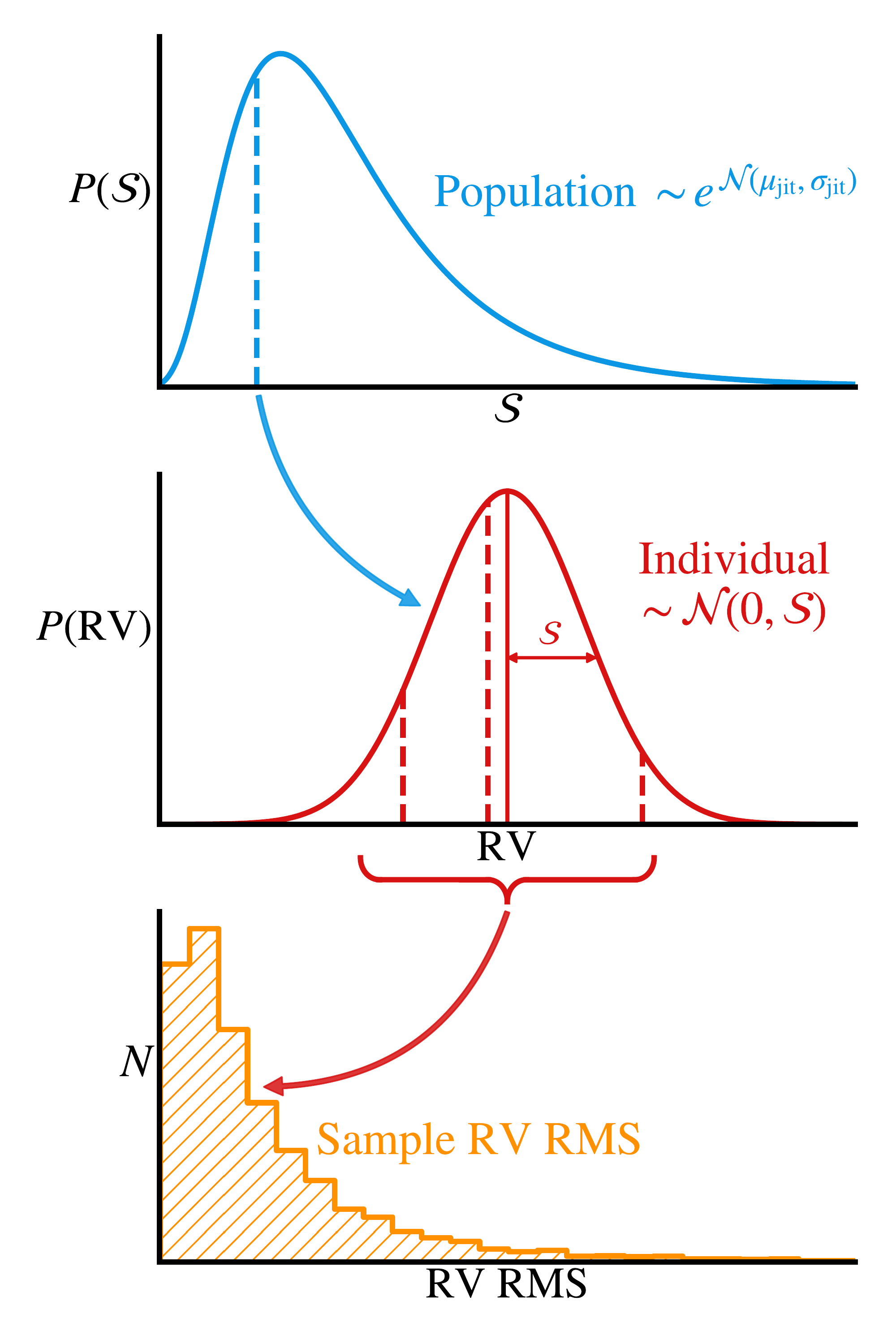}}
	    \caption{Schematic summarizing the hierarchical model we adopt for stellar jitter. For each synthetic star, a jitter value is drawn (dotted line) from the population level log-normal distribution of NIR jitter values (top, blue). This jitter value represents the standard deviation of a Gaussian distribution centered at 0, from which the RMS for an individual star is calculated using three randomly sampled RV values (center, red). This is repeated \textit{N} times to create a synthetic population of RV RMS values to mimic our observations (bottom, orange).}
	    \label{fig:syn_model}
    \end{figure}

    At every step in the chain, these two parameters are varied and a sample of 500 synthetic stars are considered. For each synthetic star, a jitter value ($\mathcal{S}$) is randomly drawn from the population-level log-normal distribution parameterized by $\mu_\mathrm{jit}$ and $\sigma_\mathrm{jit}$. This jitter value describes the characteristic amplitude of stellar jitter for that particular synthetic star and for that trial. Stellar jitter is considered to be fixed for a given star, with the ``instantaneous'' value for a given epoch drawn randomly from a normal distribution with a mean value of 0 and a standard deviation of $\mathcal{S}$: $P(j_i) = N(0, \mathcal{S}^2)$, where $j_i$ is the jitter value at epoch $i$; $j_i$ will vary from epoch to epoch and can add or subtract from measured RV values. To reflect this, synthetic jitter measurements for each epoch are drawn from a normal distribution centered at zero with a standard deviation equal to the trial jitter standard deviation $\mathcal{S}$. Three synthetic RV values are drawn from this normal distribution to reproduce the number of observations for our sample of observed stars. The synthetic RV RMS for that star is the 1$\sigma$ standard deviation of these three values. This process is repeated for all 500 synthetic stars to reproduce a mock survey like the one we have carried out. A schematic of this population analysis is illustrated in Figure \ref{fig:syn_model}.
    
    We then compare the synthetic and observed populations and calculate the likelihood of observing our RV RMS distribution for each trial $\mu_\mathrm{jit}$ and $\sigma_\mathrm{jit}$. Both the observed and model distributions are treated in the same fashion. The data and uncertainty for the observed and synthesized population are taken to be the height of each bin and the corresponding Poisson error after normalization. We compare the two distributions using the likelihood function:
    
    \begin{equation}\label{eq:likelihood}
        \mathcal{L} \propto e^{-\frac{1}{2}\sum_i{\frac{(y_i - y_{\mathrm{m},i})^2}{\sigma^2_i}}}.
    \end{equation}
    Here, $y_i$ and $y_{\mathrm{m}, i}$ are the frequencies of the observed and synthetic populations in each $i$th bin and $\sigma^2_i$ is the Poisson error for the observed distribution multiplied by a weighting factor of 0.01. We re-weight the errors to account for low number statistics, where Poisson errors are large enough that any model distributions can fit the observed distribution. Furthermore, we account for bins with zero observed systems by setting their errors to the median error of the distribution. The priors adopted for each parameter are given in Table \ref{tab:mcmc_results}. We chose broad uniform priors for each parameter since the jitter distribution at these stellar ages has not been previously measured in the near-infrared. We allow $\mu_\mathrm{jit}$ to vary between $-14.7$ and $5.7$ and $\sigma_\mathrm{jit}$ to vary between $0.0$ and $4.5$ in log-space (corresponding to bounds in linear-space of $[1, 300]$ and $[0.01, 250]$ m s$^{-1}$ for the $\mu$ and $\sigma$ of the underlying, related normal distribution). Posterior probability distributions were sampled using the affine-invariant ensemble sampler \texttt{emcee} \citep{Foreman-Mackey2013}. Since our priors are so broad, we use a total of 1000 walkers (or chains) and run each for $5 \times 10^{3}$ steps. The initial 10\% of each chain is considered as a burn-in sample and removed in the final parameter estimation. Chains are visually confirmed for stability and convergence.
    
\begin{figure*}[!ht]
        \hspace*{0.25cm}
	    \centerline{\includegraphics[width=1.05\linewidth]{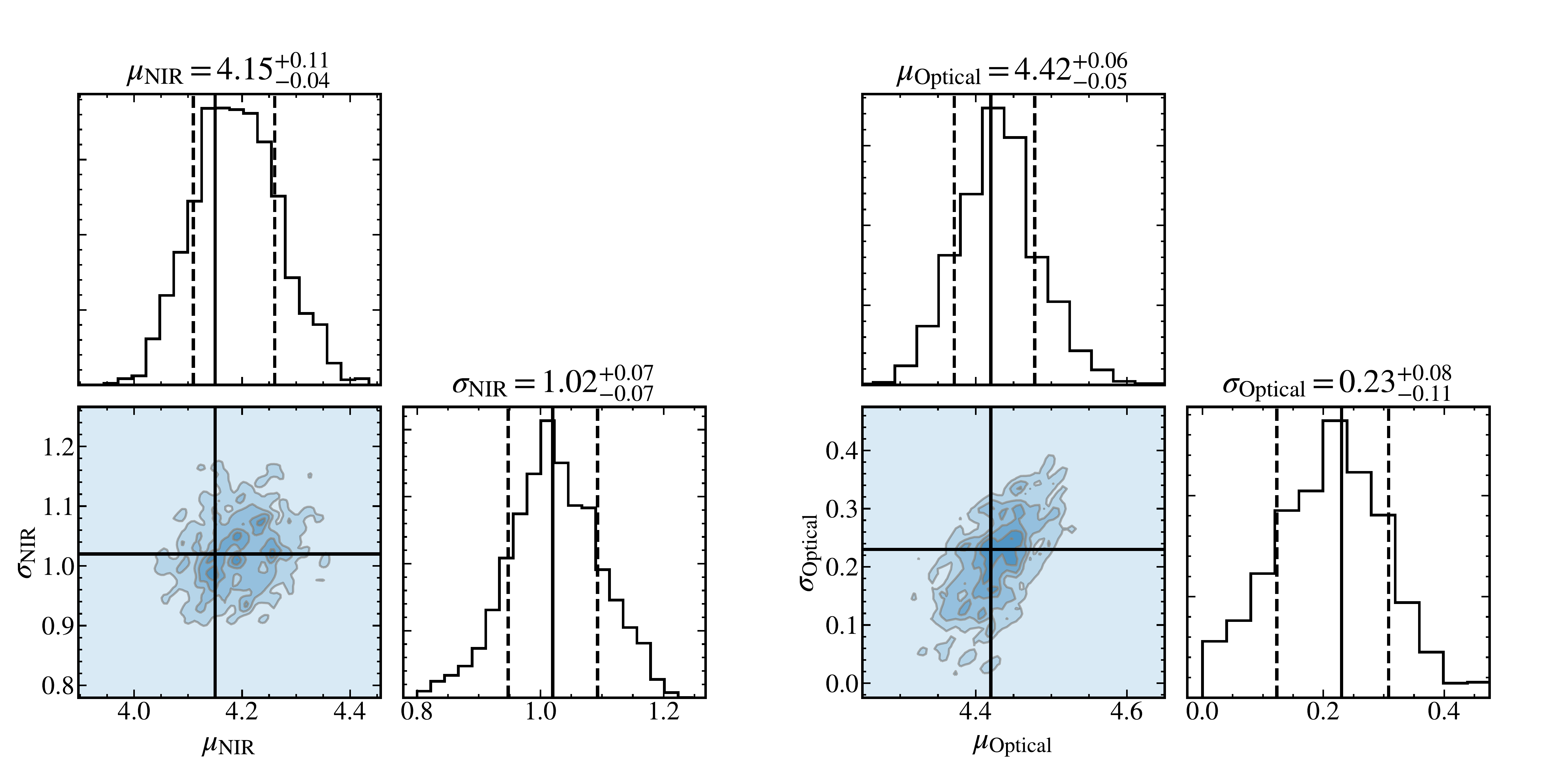}}
	    \caption{Corner plots from our MCMC fit between synthetic and observed RV RMS distributions. The comparison to our NIR observations is shown on the left and the optical compilation on the right. Contours show the covariance between the two parameters describing our population-level log-normal model for jitter, $\mu_\mathrm{jit}$ and $\sigma_\mathrm{jit}$. The modes and 68\% credible intervals of each parameter are plotted on the corresponding PDFs in solid and dashed lines, respectively.}
	    \label{fig:mcmc_jit_only_corner}
    \end{figure*}

\begin{figure*}[!ht]
	    \centerline{\includegraphics[width=1.05\linewidth]{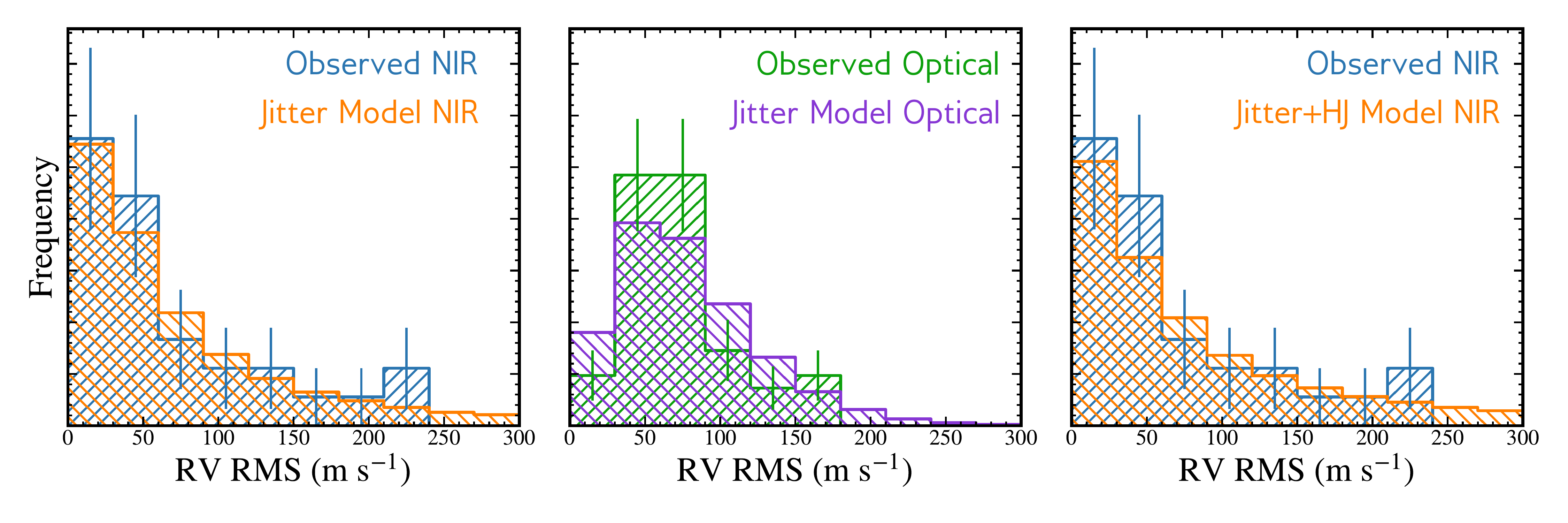}}
	    \caption{The synthesized population of RV RMS distributions for three different population-level models: the NIR sample with a jitter-only model (left), the optical sample with a jitter-only model (center), and the NIR sample with a jitter + HJs model (right).}
	    \label{fig:mcmc_diff_models}
    \end{figure*}

    Figure \ref{fig:mcmc_jit_only_corner} displays corner plots of the posterior distributions of this population synthesis fit and the best fit parameters for the NIR and optical data. $\mu_\mathrm{NIR}$ is less than $\mu_\mathrm{Optical}$, reflecting the lower characteristic value of the NIR data compared to the optical literature data. However, $\sigma_\mathrm{NIR}$ is larger than $\mu_\mathrm{Optical}$ due to a low tail to higher RV RMS values. Figure \ref{fig:mcmc_diff_models} shows the resulting synthesized populations of RV RMS distributions using the best fit parameters of the NIR and optical MCMC fits.
    
\subsection{Jointly Modelling Stellar Jitter and Hot Jupiter Frequency} \label{sec:stat_hj}
    
    To investigate all possible contributions to the measured RV RMS distribution of our survey, we include an additional HJ frequency component to our model. This is motivated by the handful of stars in our sample with unusually high RMS values ($150-200$ m s$^{-1}$), which could be caused by activity but is also consistent with signals from close-in giant planets.
    
\begin{figure}[!t]
	    \centerline{\includegraphics[width=1.21\linewidth]{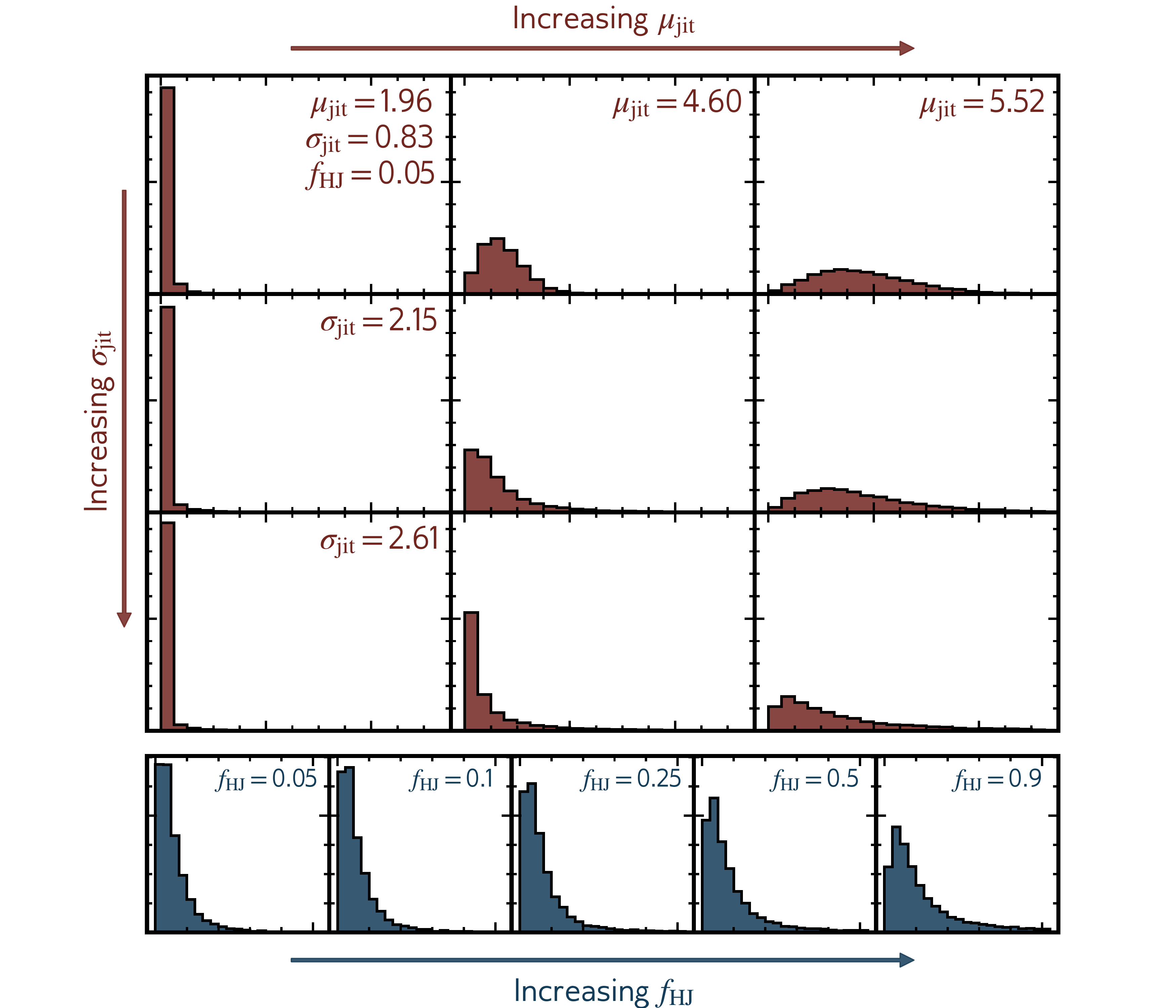}}
	    \caption{Examples of synthetic RV RMS distributions with different model parameters. In the top panels, the log-normal jitter distribution parameters $\mu_\mathrm{jit}$ and $\sigma_\mathrm{jit}$ are varied while HJ frequency is kept constant at 5\%. In the lower panels, HJ frequency is varied from 5\% to 90\% and $\mu_\mathrm{jit} = 4.16$ and $\sigma_\mathrm{jit} = 0.67$ (corresponding to 80 m s$^{-1}$ and 60 m s$^{-1}$, respectively).}
	    \label{fig:syn_lognormal}
    \end{figure}

    For each trial, an HJ frequency ($f_\mathrm{HJ}$) is randomly chosen. Since the HJ frequency at these ages is unknown, we allow $f_\mathrm{HJ}$ to take on any value between 0.0 and 1.0. Based on the HJ frequency for that trial, each synthetic star in our model sample is assigned a planet with a probability equal to the trial occurrence rate. If the star is not assigned a HJ, then the RV RMS value for that trial is calculated the same way as in the jitter only model. If the star hosts a HJ, then a stellar mass ($M_*$), planet mass ($M_p$), inclination ($i$), and orbital period ($P$) are randomly chosen while the argument of periastron ($\omega$) and eccentricity ($e$) are both set to 0, as we assume circular orbits for all HJs. The orbital period is uniformly drawn from the period range of HJs ($0.5 < P < 10$ days). The stellar host mass is drawn uniformly from the mass distribution of our sample ($0.7< M_* < 1.2$ M$_\odot$). The inclination is randomly sampled from a sin$i$ distribution by drawing $i$ uniformly from a cos$i$ distribution. Planet mass is drawn from the mass distribution of giant planets measured by \citet{Cumming2008}:

    \begin{equation} \label{eq:mass_dist}
        dN \propto M_p^{\alpha} \, d\: \mathrm{ln}M_p,
    \end{equation}
    where $\alpha = -0.31 \pm 0.2$. For a linear differential in planet mass, this is equivalent to $dN \propto M_p^{\alpha - 1} \: dM_p$. For this study we neglect uncertainties in this relation and use the mean value of alpha ($-0.31$). Using these parameters, we calculate a mean anomaly, $M(t)$, and use the Newton-Ralphson iterative method to solve Kepler's equation for the eccentric anomaly, $E(t)$,
    
    \begin{equation} \label{eq:keplers}
    M(t) = \frac{2\pi}{P}\left(t - t_0\right) = E(t) - e\, \mathrm{sin} \, E(t),
    \end{equation}
    where $t$ is time and $t_0$ is the time of pericenter. Using the eccentric anomaly for our model system, we calculate the true anomaly $f(t)$ from
    
    \begin{equation} \label{eq:true_anom}
    \mathrm{tan}\left(\frac{f(t)}{2}\right) = \sqrt{\frac{1+e}{1-e}}\mathrm{tan}\left(\frac{E(t)}{2}\right).
    \end{equation}
    We then compute the RV semi-amplitude,
    
    \begin{equation}\label{eq:K}
        K = \frac{2\pi \: a \: \mathrm{sin}i}{P \sqrt{1 - e^2}},
    \end{equation}
    where $a$ is the semi-major axis and is calculated using Kepler's Third Law with the random period $P$ and stellar mass $M_*$. Then, over a large time range, RVs are calculated using the following relation:
    
    \begin{equation}\label{eq:rv_vels}
        V(t) = K \Big( \mathrm{cos}\bigl(\omega + f(t) \bigr) + e \, \mathrm{cos} \, \omega \Big).
    \end{equation}
    Three velocities are randomly drawn from the Keplerian curve to simulate the typical number of epochs in our subsample and the semi-random observing cadence of our HPF observations. RV jitter for each epoch is calculated in the same fashion as in the non-planet case, and the adopted RV for that epoch is a linear sum of the jitter value and the dynamically-induced RV. The RV RMS is then calculated from these final three velocities. As before, this is carried out for all 500 synthetic stars in our sample for that particular trial. Figure \ref{fig:syn_lognormal} demonstrates how the resulting RV RMS distribution changes a function of each of these model parameters ($\mu_\mathrm{jit}$, $\sigma_\mathrm{jit}$, and $f_\mathrm{HJ}$).
    
    To explore the full range of potential results, we first fit our observed NIR distribution to a model with only planets and no jitter. We report the best fit parameters in Table \ref{tab:mcmc_results}. The data are well described by a planet-only model, but this requires an improbably high HJ occurrence rate of 95\%.
    
\begin{figure}[!ht]
	    \centerline{\includegraphics[width=1.15\linewidth]{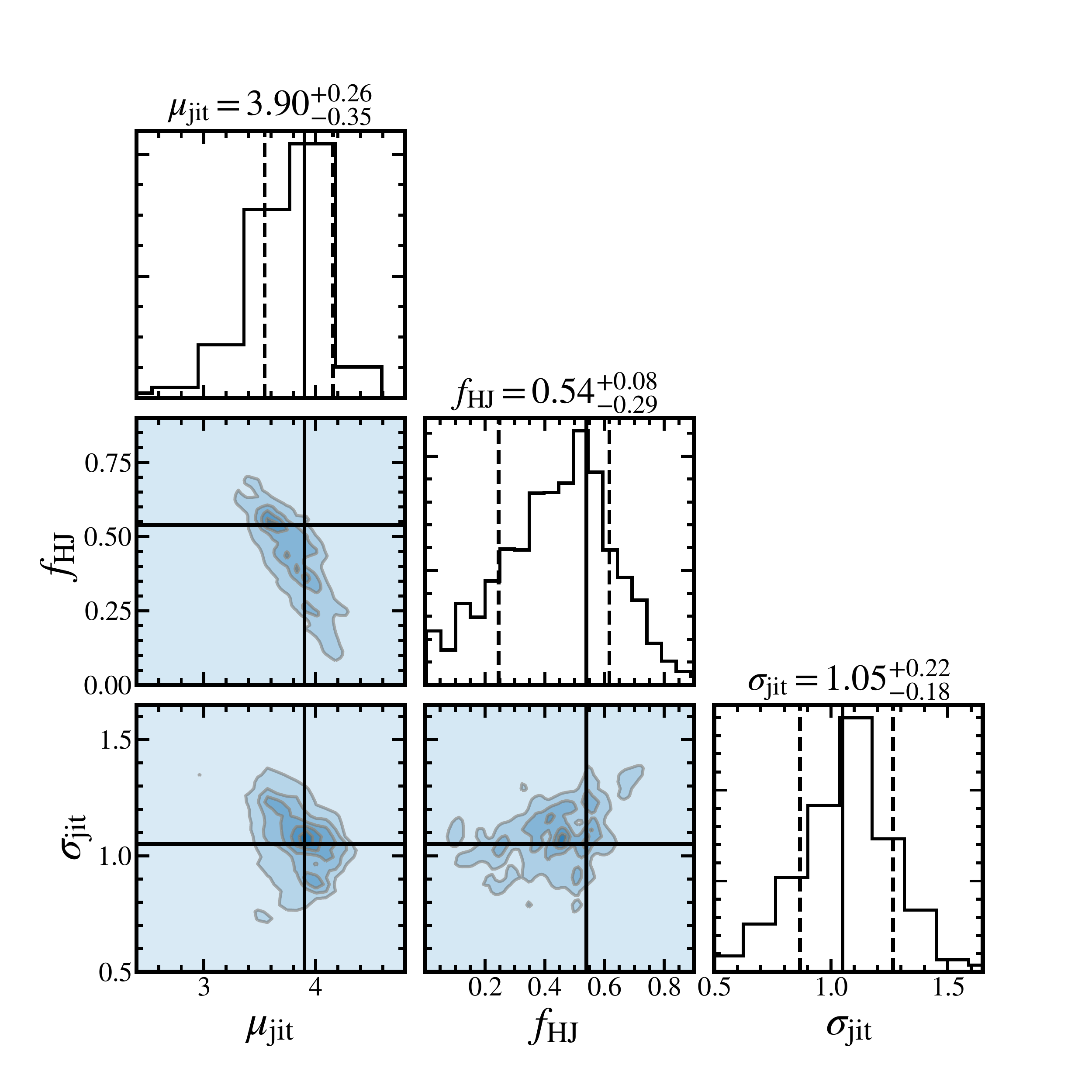}}
	    \caption{Corner plot from our MCMC results assuming a two-component model of stellar jitter and HJ occurrence rate. Contours show the covariance between the three parameters describing our model: $\mu_\mathrm{jit}$, $\sigma_\mathrm{jit}$, and $f_\mathrm{HJ}$. The modes and 68\% credible intervals of each parameter are plotted on the corresponding corresponding marginalized posteriors in solid and dotted lines, respectively. The HJ frequency parameter prefers moderately high values of 0.54, but remains unconstrained with a 95\% credible interval of 0.0 -- 0.75.}
	    \label{fig:emcee_synthesis}
    \end{figure}

    To assess the performance of a two-component planet and jitter model, we jointly fit our data with the combined jitter and HJ frequency model. The best fit parameters, the resulting synthetic RV RMS distribution, and the observed RV RMS distribution are plotted in Figure \ref{fig:mcmc_diff_models}. The corner plot of the posterior distribution of each parameter is shown in Figure \ref{fig:emcee_synthesis}. The best fit and $\chi^2$ values are reported in Table \ref{tab:mcmc_results}. Our observed RV RMS values are most consistent with a synthesized population with individual jitter values drawn from a log-normal jitter distribution with $\mu_\mathrm{jit} = 3.90^{+0.26}_{-0.35}$ and $\sigma_\mathrm{jit} = 1.05^{+0.22}_{-0.18}$. The HJ frequency favors a high value, with a mode at $f_\mathrm{HJ, Mode} = 0.54^{+0.08}_{-0.29}$. However, it remains poorly constrained with an upper limit of 0.75 at the 95\% confidence level. This suggests that some of the stars exhibiting high-RMS values can be explained by HJs, or could be attributed to anomalously active stars that are not well represented by a log-normal population-level model. This hypothesis can be tested for these ``objects of interest'' with follow-up multi-wavelength precision RVs.
    
    \setlength{\tabcolsep}{3.4pt}
    \renewcommand{\arraystretch}{1.3}

    \begin{table}[!tp]
    \scriptsize
    \centering
    \caption{MCMC Priors and Best Fit Parameters}
        \begin{tabular}{c | c c c}
        \hline\hline 
        Parameter & $\mu_\mathrm{jit}$ & $\sigma_\mathrm{jit}$ & $f_\mathrm{HJ, Mode}$ \\
        \hline
        Prior & $\mathcal{U} [-14.7, 5.7]$ & $\mathcal{U} [0.0, 4.5]$ & $\mathcal{U} [0.0, 1.0]$ \\
        \hline\hline 
        Synthesis Model & $\mu_\mathrm{jit}$ & $\sigma_\mathrm{jit}$ & $f_\mathrm{HJ, Mode}$ \\ [0.5ex]
        \hline
        Optical Jitter Only & $4.42^{+0.06}_{-0.05}$ & $0.23^{+0.08}_{-0.11}$ & $\cdots$ \\
        
        NIR Jitter Only & $4.15^{-0.04}_{+0.11}$ & $1.02^{+0.07}_{-0.07}$ & $\cdots$ \\
        
        NIR HJs Only & $\cdots$ & $\cdots$ & $0.95^{+0.05}_{-0.13}$ \\
        
        NIR Jitter + HJs & $3.90^{+0.26}_{-0.35}$ & $1.05^{+0.22}_{-0.18}$ & $0.54^{+0.08}_{-0.29}$ \\
        [1ex]
        \hline
        \end{tabular}
    \label{tab:mcmc_results}
    \end{table}

    \begin{figure*}[!ht]
	    \centerline{\includegraphics[width=1.025\linewidth]{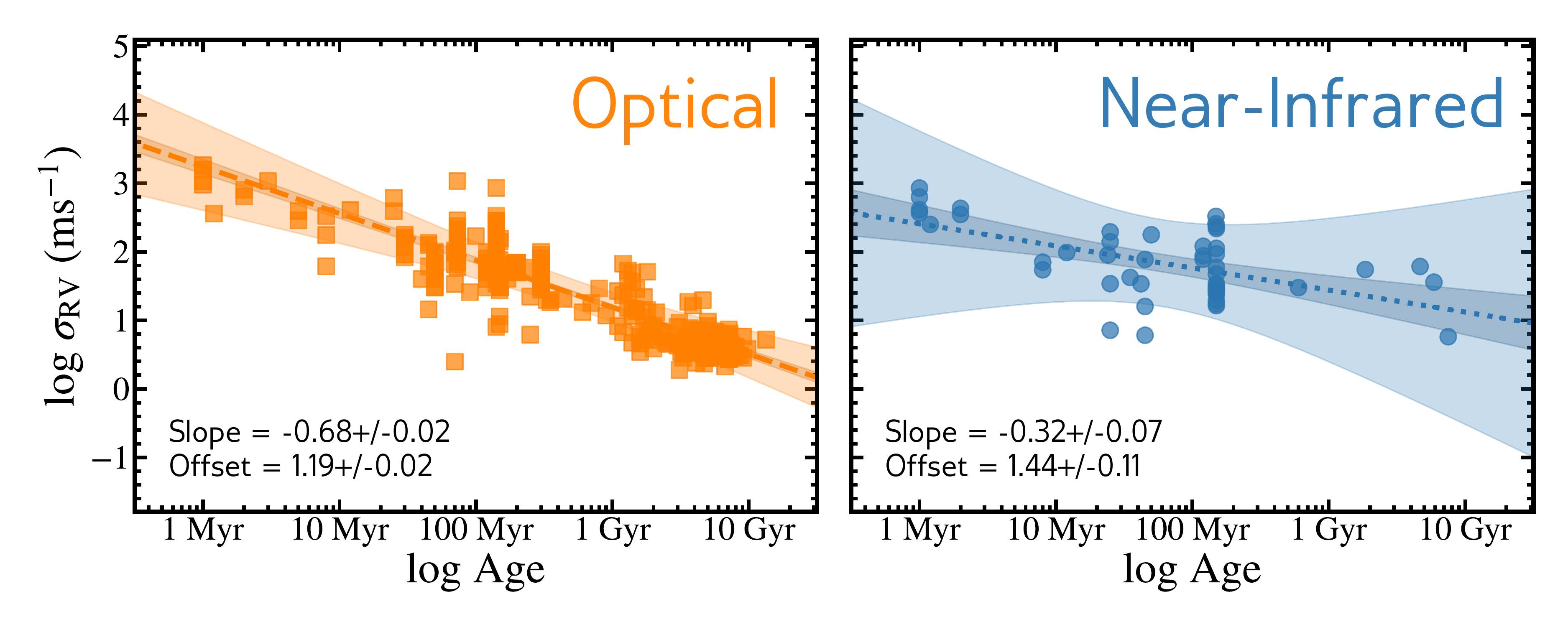}}
	    \caption{The relationship between optical (orange, left) and NIR (blue, right) jitter and stellar age. The orange dashed and blue dotted lines are linear fits to the data and the dark and light shaded regions are the 95\% confidence and prediction bands for each fit, respectively. Note that the confidence band is the area that has a 95\% chance of containing the true regression line and is a reflection of the errors on the fitted parameters. The prediction band is an estimate of where 95\% of all points are expected to fall and is therefore wider than a confidence band. References for measured RV RMS values and ages are reported in Table \ref{tab:log_relation}. The decay of jitter with stellar age in both the optical and NIR is well-modelled with a power law.}
	    \label{fig:log_relation}
    \end{figure*}

\section{\textbf{Discussion}}
\label{sec:Dicussion}
    
    \begin{table}[!tp]
    \scriptsize
    \setlength{\tabcolsep}{2.5pt}
    \renewcommand{\arraystretch}{1.15}
    \centering
    \caption{Sources for optical and near-infrared stellar jitter sample }
        \begin{tabular}{c | c c}
        \hline\hline 
        Optical/NIR & $N$ & Reference \\ [0.5ex]
        \hline
        Optical & 43 & \citet{Paulson2006} \\
        Optical & 7 & \citet{Crockett2012} \\
        Optical & 12 & \citet{Lagrange2013} \\
        Optical & 62 & \citet{Bailey2018} \\
        Optical & 11 & \citet{Brems2019} \\
        Optical & 46 & \citet{Grandjean2020} \\
        Optical & 153 & \citet{Brewer2016, Luhn2020} \\
        NIR & 2 & \citet{Figueira2010b} \\  
        NIR & 10 & \citet{Crockett2012} \\
        NIR & 2 & \citet{Bailey2012} \\
        NIR & 4 & \citet{Gagne2016} \\
        NIR & 29 & This study \\ [1ex]
        \hline
        \end{tabular}

    \label{tab:log_relation}
    \end{table}

    RV variability due to starspot coverage is expected to diminish over a star's lifetime as the star spins down and its magnetic field weakens. To compare how jitter levels among Sun-like stars decay in the optical and near-infrared, we compile a sample of 334 RV RMS measurements in the optical. To ensure a fair comparison with our sample in the near-infrared, we isolate stars with G and K spectral types and $v$sin$i$ $< 30.0$ km s$^{-1}$. This literature sample is an extension of the stars assembled in Section \ref{sec:optical_jitter} but for a broader age range. The stars in this sample are drawn from precision optical RV surveys with reliable age estimates and have properties consistent with our survey but with ages ranging from 1 Myr--10 Gyr (see Table \ref{tab:log_relation}). Stars with known giant planets have been removed. A power law decay of stellar jitter with time provides a good fit to the data. A linear fit to the optical RV RMS sample gives:
    
    \begin{equation}
        \mathrm{log}\left(\mathrm{RMS}_\mathrm{Optical, RV}\right) = -0.68 \times \mathrm{log}(\tau) + 1.19,
    \end{equation}
    where $\mathrm{RMS}$ is the stellar jitter in m s$^{-1}$ and $\tau$ is the age in Gyr. The uncertainty in the slope and offset are 0.02 Gyr m$^{-1}$ s$^{-1}$ and 0.02 m s$^{-1}$, respectively, with modest covariance between the two parameters. Figure \ref{fig:log_relation} displays the observations and best-fit power law decay.
    
    With our new sample of 29 intermediate-age stars we can also investigate for the first time whether this relationship holds in the NIR. We compile a literature sample of precision NIR RV RMS values with the same stellar characteristics as our optical sample (G and K spectral types and $v$sin$i$ $<$ 30 km s$^{-1}$); see Table \ref{tab:log_relation} for a detailed breakdown of the sources that went into creating this sample. The subsample of 29 stars from our survey increases the total number of stars with NIR RV RMS measurements in this range of rotational velocities and spectral types by 150\%. As with the optical data, we apply a linear fit to the NIR observations in log-log space and find the following best-fit decay curve:
    
    \begin{equation}
        \mathrm{log}\left(\mathrm{RMS}_\mathrm{NIR, RV}\right) = -0.32 \times \mathrm{log}(\tau) + 1.44.
    \end{equation}
    
    Uncertainties in the slope and jitter offset are 0.07 Gyr m$^{-1}$ s$^{-1}$ and 0.11 m s$^{-1}$, respectively. We note that the slope of this linear fit is shallower than the optical sample at the 4.5-sigma level, indicating that the near-infrared activity level may decay more slowly than at optical wavelengths.
    
    RV scatter obeys a similar logarithmic relationship with stellar age in both the optical and NIR. This trend is shallower in the NIR such that at younger ages (less than several hundred Myr), RV RMS values are lower in NIR than in the optical, while at older ages, this tendency appears to be reversed. This may reflect (historically) poorer RV precision of infrared spectrographs, which would increase RV RMS levels for old, quiet stars compared to measurements at optical wavelengths. That is, intrinsic NIR instrumental errors may establish a systematic RV floor for the sample of old stars in Figure \ref{fig:log_relation}. Regardless, the large scatter in the NIR relation---itself likely driven by the comparably small number of very young and old stars with NIR jitter measurements---makes a more meaningful assessment difficult at this time. It is nevertheless clear that jitter decays in a qualitatively similar manner in the NIR as has been seen in the optical. Although the scatter is large in these relations, NIR jitter is also systematically reduced by a factor of $\sim$2 at young and intermediate ages ($\lesssim$1 Gyr) which reflects the reduced starspot-to-photosphere contrast at longer wavelengths.  This is in line with conclusions from previous comparisons of individual stars with both optical and NIR RVs \citep{Crockett2012, Bailey2012, Gagne2016}.

\section{\textbf{Summary}} \label{sec:Summary}
    The Epoch of Giant Planet Migration planet search program is a precise RV survey of intermediate-age ($\sim$20--200 Myr) G and K dwarfs with the Habitable--Zone Planet Finder spectrograph (HPF) at McDonald Observatory's Hobby-Eberly Telescope (HET). The goal of our survey is to determine the timescale and dominant physical mechanism of giant planet migration by measuring the frequency of giant planets interior to the water ice line at intermediate ages and to eventually compare this result to existing field measurements over the same ranges of planet mass, orbital separation, and stellar mass.
    
    As part of this effort, we created a custom HPF RV extraction pipeline and have demonstrated $<$2 m s$^{-1}$ binned precision on the K2 RV standard star HD 3765.
    
    Using a subsample of 29 stars from our survey with at least three epochs of RV measurements and a comparison sample at optical wavelengths, we attempt to constrain the general properties of the underlying distribution of stellar jitter. We model the distribution of stellar jitter for our subsample as a log-normal distribution at the population level and find best-fit parameters of $\mu_\mathrm{jit} = 4.15^{+0.11}_{-0.04}$ and $\sigma_\mathrm{jit} = 1.02^{+0.07}_{-0.07}$ for the NIR sample. A comparison sample of RV RMS values at optical wavelengths compiled from the literature yield best-fit parameters of $\mu_\mathrm{jit} = 4.42^{+0.06}_{-0.05}$ and $\sigma_\mathrm{jit} = 0.23^{+0.08}_{-0.11}$. We also jointly constrain the contribution of stellar jitter and HJ frequency on the observed RV measurements to investigate how planets might contribute to our observed RV RMS distribution. We find that this joint model gives best-fit parameters of $\mu_\mathrm{jit} = 3.90^{+0.26}_{-0.35}$, $\sigma_\mathrm{jit} = 1.05^{+0.22}_{-0.18}$, and $f_\mathrm{HJ, Mode} = 0.54^{+0.08}_{-0.29}$. This suggests that our high-RMS targets could be caused by close-in planets, or perhaps from a population-level jitter model with a broader tail to high values.
    
    To explore how jitter levels evolve over time, we constrain the power law relationship between RV scatter and stellar age in the optical and assess a similar trend in the NIR. We find evidence that the decay of jitter in the NIR is shallower than in the optical, although the scatter is large and it is unclear if differences with the optical jitter decay reflect stellar activity or instead are caused by small number statistics, selection biases, or perhaps instrument precision. A larger sample of RV RMS measurements, both at very young and old ages, would help to refine this fit. Regardless, we measure a median NIR RV RMS of 34.2 m s$^{-1}$ at intermediate ages, which is reduced by nearly a factor of two from the optical RV jitter values for a sample of stars with similar ages, rotational velocities, and spectral types. This finding aligns well with previous observations and points toward a bright future of exoplanet science for active stars at NIR wavelengths.
    
    \acknowledgments
    We thank Benjamin Tofflemire, Daniel Krolikowski, Andrew Vanderburg, Aaron Rizzuto, Adam Kraus, and Michael Gully-Santiago for their helpful insights on precision RV reduction and the impact of starspots on jitter. 
    
    Q.H.T. and B.P.B. acknowledge the support from a NASA FINESST grant (80NSSC20K1554). B.P.B. acknowledges support from the National Science Foundation grant AST-1909209.
    
    These results are based on observations obtained with the Habitable--Zone Planet Finder Spectrograph on the 9.2-meter Hobby-Eberly Telescope at McDonald Observatory. The authors thank the resident astronomers and telescope operators at the HET for obtaining these observations.
    
    This work was partially supported by funding from the Center for Exoplanets and Habitable Worlds. The Center for Exoplanets and Habitable Worlds is supported by the Pennsylvania State University, the Eberly College of Science, and the Pennsylvania Space Grant Consortium. This work was supported by NASA Headquarters under the NASA Earth and Space Science Fellowship Program through grants NNX16AO28H. SM, JPN and GS acknowledge support from NSF grants AST-1006676, AST-1126413, AST-1310885, AST-1517592, AST-1310875, AST-1907622, ATI-2009889, the NASA Astrobiology Institute (NAI; NNA09DA76A), and PSARC in our pursuit of precision radial velocities in the NIR. We acknowledge support from the Heising-Simons Foundation via grant 2017-0494 and 2019-1177.
    
    The Hobby-Eberly Telescope is a joint project of the University of Texas at Austin, the Pennsylvania State University, Ludwig-Maximilians-Universität München, and Georg-August Universität Gottingen. The HET is named in honor of its principal benefactors, William P. Hobby and Robert E. Eberly. The HET collaboration acknowledges the support and resources from the Texas Advanced Computing Center. Computations for this research were also performed on the Pennsylvania State University’s Institute for Computational and Data Sciences Advanced CyberInfrastructure (ICDS-ACI, now known as Roar), including the CyberLAMP cluster supported by NSF grant MRI1626251.
    
    \textit{Facilities}: HET (HPF).
    
    \textit{Software}: \texttt{HxRGproc} \citep{Ninan2018},
    \texttt{GNU parallel} \citep{Tange2011a}, \texttt{astroquery} \citep{Ginsburg2019}, \texttt{astropy} \citep{Astropy2018}, \texttt{barycorrpy} \citep{Kanodia2018}, \texttt{matplotlib} \cite{Hunter4160265}, \texttt{numpy} \citep{vanderWalt2011}, \texttt{pandas} \citep{mckinney-proc-scipy-2010}, \texttt{SERVAL} \citep{Zechmeister2018}, \texttt{PyMC3} \citep{Salvatier2016}, \texttt{emcee} \citep{Foreman-Mackey2013}, \texttt{scipy} \citep{Virtanen2020}.

\appendix

\section{Properties of the Subsample of Stars Analyzed in this Study}\label{sec:appendix}
Table \ref{tab:sample_info} lists the properties and RV RMS values of the subsample of 29 stars used in this analysis based on the first 14 months of HPF observations from this survey. See Section \ref{sec:nir_jitter} for details.
 
\begin{deluxetable*}{ccccccccccccc}
\tablecaption{Properties of the Subsample of 29 Stars Analyzed in this Study \label{tab:sample_info}}
\tablewidth{475pt}
\tabletypesize{\scriptsize}
\tablehead{
\colhead{Target Name} & \colhead{RV RMS} & \colhead{$N$\tablenotemark{\scriptsize{a}}} & \colhead{$\bar{\sigma}_\mathrm{Binned}$\tablenotemark{\scriptsize{b}}} & \colhead{$\Delta t$\tablenotemark{\scriptsize{c}}} & 
\colhead{$J$\tablenotemark{\scriptsize{d}}} & \colhead{SpT} & \colhead{SpT} & \colhead{Lit. $v$sin$i$} & \colhead{$v$sin$i$} & \colhead{Meas. $v$sin$i$\tablenotemark{\scriptsize{e}}} & \colhead{Age} & \colhead{Age} \\ 
\colhead{} & \colhead{(m s$^{-1}$)} & \colhead{} & \colhead{(m s$^{-1}$)} & \colhead{(days)} & \colhead{(mag)} & \colhead{} & \colhead{Ref.}& \colhead{(km s$^{-1}$)} & \colhead{Ref.} & \colhead{(km s$^{-1}$)} & \colhead{(Myr)} & \colhead{Ref.}
}
\startdata
2MASS J19224278--0515536 & 177.2 & 3 & 71.2 & 275 & 9.9 & K5 & 1 & 9$\pm$1 & 2 & 30$\pm$8 & 50 & 3 \\
BD-03 5579 & 77.3 & 3 & 35.2 & 25 & 8.6 & K4 & 4 & 13$\pm$1 & 5 & 13$\pm$3 & 45 & 6 \\
BD+17 455 & 6.06 & 3 & 10.6 & 101 & 7.6 & K0 & 7 & \nodata & \nodata & 4$\pm$1 & 45 & 8 \\
BD+20 1790 & 24.2 & 7 & 25.9 & 22 & 7.6 & K5 & 9 & 16$\pm$3 & 10 & 16$\pm$2 & 150 & 8 \\
BD+21 418 & 47.1 & 5 & 17.2 & 458 & 7.3 & G5 & 11 & 8$\pm$1 & 12 & 11$\pm$1 & 150 & 8 \\
BD+21 418B & 16.4 & 6 & 11.4 & 384 & 8.4 & K6 & 13 & 6$\pm$2 & 14 & 7$\pm$1 & 150 & 8 \\
BD+25 430 & 87.6 & 4 & 41.7 & 424 & 8.2 & G0 & 15 & \nodata & \nodata & 6$\pm$1 & 25 & 8 \\
BD+26 592 & 87.6 & 3 & 41.7 & 144 & 9.0 & G0 & 7 & 5$\pm$2 & 16 & 6$\pm$1 & 120 & 17 \\
BD+41 4749 & 25.4 & 4 & 10.4 & 97 & 7.6 & G0 & 18 & 4$\pm$1 & 12 & 5$\pm$1 & 150 & 8 \\
BD+49 646 & 231.1 & 3 & 55.5 & 364 & 8.0 & K2 & 5 & 26$\pm$1 & 5 & 29$\pm$5 & 150 & 19 \\
Cl* Melotte 22 DH 875 & 79.0 & 3 & 100 & 357 & 11.5 & K5 & 20 & \nodata & \nodata & 26$\pm$5 & 120 & 17 \\
HD 14082B & 34.3 & 3 & 6.6 & 17 & 6.6 & G2 & 11 & \nodata & \nodata & 9$\pm$2 & 25 & 8 \\
HD 16760 & 59.3 & 3 & 11.3 & 26 & 7.4 & G2 & 21 & 2$\pm$2 & 12 & 2$\pm$1 & 150 & 8 \\
HD 16760B & 29.1 & 6 & 24.9 & 4 & 8.4 & K2 & 21 & 9$\pm$5 & 5 & 7$\pm$1 & 150 & 8 \\
HD 189285 & 32.5 & 3 & 50.3 & 92 & 8.3 & G5 & 22 & 11$\pm$2 & 5 & 10$\pm$2 & 150 & 12 \\
HD 21845 & 17.4 & 4 & 19.8 & 354 & 6.8 & K2 & 18 & 9$\pm$1 & 12 & 10$\pm$1 & 150 & 8 \\
HD 221239 & 27.8 & 3 & 11.2 & 14 & 6.7 & K2 & 23 & \nodata & \nodata & 3$\pm$1 & 150 & 19 \\
HD 236717 & 19.6 & 4 & 13.4 & 231 & 7.2 & K0 & 24 & \nodata & \nodata & 4$\pm$1 & 150 & 19 \\
HD 24681 & 112.2 & 3 & 9.3 & 408 & 7.7 & G5 & 22 & \nodata & \nodata & 18$\pm$3 & 150 & 19 \\
HD 287167 & 7.2 & 3 & 24.0 & 21 & 8.3 & K0 & 15 & \nodata & \nodata & 2$\pm$2 & 25 & 19 \\
HD 48370 & 34.2 & 3 & 11.7 & 37 & 6.7 & K0 & 22 & \nodata & \nodata & 10$\pm$2 & 42 & 25 \\
HS Psc & 218.0 & 8 & 81.8 & 250 & 8.4 & K5 & 26 & 10$\pm$2 & 14 & 22$\pm$8 & 150 & 8 \\
IS Eri & 34.1 & 3 & 8.1 & 349 & 7.2 & G8 & 22 & 6 & 27 & 8$\pm$1 & 150 & 8 \\
LP 745-70 & 28.5 & 3 & 21.0 & 45 & 8.4 & K9 & 23 & \nodata & \nodata & 10$\pm$2 & 150 & 8 \\
NX Aqr & 42.3 & 3 & 14.9 & 97 & 6.4 & G5 & 4 & 4 & 28 & 4$\pm$1 & 35 & 29 \\
PW And & 93.2 & 3 & 47.2 & 191 & 7.0 & K2 & 30 & 23$\pm$1 & 31 & 27$\pm$4 & 150 & 19 \\
StKM 1-382 & 15.9 & 3 & 15.7 & 112 & 9.3 & K4 & 26 & \nodata & \nodata & 4$\pm$1 & 45 & 8 \\
V1274 Tau & 195.3 & 3 & 87.2 & 110 & 8.6 & K5 & 15 & \nodata & \nodata & 9$\pm$2 & 25 & 19 \\
V370 Tau & 120.2 & 3 & 72.1 & 79 & 11.0 & K3 & 32 & \nodata & \nodata & 6$\pm$2 & 120 & 17 \\
\enddata
\tablecomments{RV RMS values are based on a minimum of three RV measurements within the first 14 months of observations from the larger four-year program of over 100 targets.}
\tablenotetext{a}{Number of epochs used for RV RMS calculation.}
\tablenotetext{b}{Average RV measurement error of all target epochs, each of which are binned from 3 contiguous observations.}
\tablenotetext{c}{Time between first epoch and last epoch.}
\tablenotetext{d}{\citet{Cutri2003}}
\tablenotetext{e}{Rotational velocity measured from HPF observations using \texttt{iSpec} \citep{Blanco-Cuaresma2019}. See Section \ref{sec:vsini} for more details.}
\tablerefs{(1) \citet{Riaz2006}, (2) \citet{Malo2013}, (3) \citet{Bell2015}, (4) \citet{Torres2006}, (5) \citet{Frasca2018}, (6) \citet{Ramirez-Preciado2018}, (7) \citet{Roeser1988}, (8) \citet{Gagne2018a}, (9) \citet{Reid2004}, (10) \citet{White2007}, (11) \citet{Egret1992}, (12) \citet{McCarthy2014}, (13) \citet{Zuckerman2004}, (14) \citet{Schlieder2010}, (15) \citet{Nesterov1995}, (16) \citet{Mermilliod2009}, (17) \citet{Stauffer2007} (18) \citet{Malo2013}, (19) \citet{Gagne2018b}, (20) \citet{Findeisen2010}, (21) \citet{Abt1988}, (22) \citet{Houk1999}, (23) \citet{Gray2003}, (24) \citet{Yoss1961}, (25) \citet{Elliott2016}, (26) \citet{Stephenson1986}, (27) \citet{Desidera2004}, (28) \citet{Luck2017}, (29) \citet{Zuckerman2013}, (30) \citet{Christian2001}, (31) \citet{Lopez-Santiago2003}, (32) \citet{Haro1982}.}
\end{deluxetable*}

%\bibliography{survey}{}
%\bibliographystyle{aasjournal}

\end{document}